\documentclass[reprint,aps,prl,showkeys,showpacs,superscriptaddress,twocolumn,amsmath,amssymb,longbibliography]
{revtex4-1} 
\usepackage{graphics}
\usepackage{bm}
\usepackage{subfigure}
\usepackage{epsfig}
\usepackage[bookmarks,colorlinks,citecolor=blue,linkcolor=blue]{hyperref}

\usepackage{natbib}
\usepackage{siunitx}
\usepackage{soul}
\usepackage{ulem} 
\usepackage{array}
\usepackage{multirow}
\usepackage{amsmath}
\usepackage{esvect}
\usepackage{bm}

\renewcommand{\vec}[1]{\bm{#1}}

\DeclareSIUnit\Molar{\textsc{m}}

\usepackage{xcolor}

\usepackage{graphicx}	
\begin{document}
\title{Scalability of Graph Neural Network in Accurate Prediction of Frictional Contact Network in Suspensions}
\author{Armin Aminimajd}
\affiliation{Department of Macromolecular Science and Engineering, Case Western Reserve University, Cleveland, OH, 10040, USA}
\author{Joao Maia}
\affiliation{Department of Macromolecular Science and Engineering, Case Western Reserve University, Cleveland, OH, 10040, USA}
\author{Abhinendra Singh}
\email{abhinendra.singh@case.edu}
\affiliation{Department of Macromolecular Science and Engineering, Case Western Reserve University, Cleveland, OH, 10040, USA}
\date{\today} 
\begin{abstract}
Dense suspensions often exhibit shear thickening, characterized by a dramatic increase in viscosity under large external forcing. This behavior has recently been linked to the formation of a system-spanning frictional contact network (FCN), which contributes to increased resistance during deformation. However, identifying these frictional contacts poses experimental challenges and is computationally expensive. This study introduces a Graph Neural Network (GNN) model designed to accurately predict FCNs in two dimensional simulations of dense shear thickening suspensions.
The results demonstrate the robustness and scalability of  the GNN model across various stress levels $(\sigma)$, packing fractions$(\phi)$, system sizes, particle size ratios$(\Delta)$, and amount of smaller particles. The model is further able to predict both the occurrence and structure of the FCN. The presented model is accurate and interpolates and extrapolates to conditions far from its control parameters. This machine learning approach provides an accurate, lower cost, and faster predictions of suspension properties compared to conventional methods, while it is trained using only small systems. Ultimately, the findings in this study pave the way for predicting frictional contact networks in real-life large-scale polydisperse suspensions, for which theoretical models are largely limited owing to computational challenges.
\end{abstract}
\maketitle
\section{Introduction}
Dense particulate suspensions are ubiquitous in natural, human health, and industrial settings, with examples ranging from mud to blood to paint and cement~\cite{Coussot_1997, Jerolmack_2019, Galdi_2008, Bridgwater_1993}.
Under external deformation, they exhibit diverse non-Newtonian complex rheological behaviors, such as yielding, normal stress differences, shear-thinning, shear thickening, and jamming~\cite{Morris_2020, Denn_2018, Singh_2023}.
Shear thickening (ST) is a non-linear phenomenon where the viscosity $\eta$ increases continuously (CST) or discontinuously (DST) with increasing shear rate $\dot{\gamma}$ at a given volume fraction $\phi$~\cite{Mewis_2011}.
A vast body of research has linked non-Newtonian rheology in dense suspensions to constraints on the relative motion that stabilizes the force and contact network under applied deformation~\cite{Morris_2020, Singh_2020, Mari_2014, Boromand_2018, Pradeep_2021, Sedes_2020,Sedes_2022, Naald_2024, Chen_2022, Rathee_2021,Amico_2025}.
%
\textcolor{black}{
As such, many recent studies have used network science techniques to characterize and analyze the properties of these
networks in suspensions ~\cite{Thomas_2018, Gameiro_2020, thomas2020investigating, Nabizadeh_2022, Sedes_2022, Edens_2019, Edens_2021, Naald_2024} as well as in colloidal gels~\cite{smith2024topological, gupta2024next, whitaker2019colloidal, bindgen2020connecting, Nabizadeh_2024} and correlate them with the
resultant rheology.}
%
Historically, the mesoscale network features have been linked to the emergence of rigidity~\cite{Silke_2016}, sound propagation~\cite{Owens_2011}, and non-locality~\cite{Thomas_2019} for both frictionless and frictional particle packings in dry granular materials. 
Therefore, accurate and easy identification of contact networks in amorphous materials, specifically in flowing dense suspensions, is highly beneficial for understanding and developing a statistical physics framework for rheological responses.

However, identifying the frictional contact network (FCN) in most particulate suspensions remains challenging owing to the limitations of the experimental methods and the complex nature of particulate systems. 
State-of-the-art experimental efforts are limited by the protocol to freeze the sheared microstructure, along with the cost and toxicity of the chemicals used to generate the model suspension system~\cite{Pradeep_2020,Pradeep_2021,Nabizadeh_2024}.
%
%
These frictional contacts are readily accessible in discrete particle simulations; however, traditional methods remain computationally expensive, making large system sizes computationally intractable, despite recent advances in computational power.
The two fundamental questions that remain unanswered are:
(i) How can frictional contacts under applied stress be predicted, \textit{i.e.}, is it possible to predict which particles will be in contact?
(ii) Can the structure of the frictional contact network for various solid concentrations, system sizes, particle size dispersities, and applied stresses be predicted?

Recently, a more promising avenue has been the use of machine-learning (ML) techniques~\cite{Mandal_2022, Li_2023_The_Prediction, ashwin2022deep, rossi2022identification, binel2023online, galloway2022relationships, Ferguson_2017,Cubuk_2015,Carleo_2019}. 
In soft matter systems, ML has been used to predict wall penetration by particles in coarse-grained simulations using the random forest method~\cite{barcelos2022supervised}, prediction of drag forces~\cite{he2019supervised}, particle stress development using physics-informed neural network~\cite{howard2023machine}, and detection of hidden correlation in particle size distribution and mechanical behavior~\cite{gonzalez2022use}.
Among the diverse ML techniques, Graph Neural Networks (GNN) have exhibited superior performance compared to traditional methods, mainly owing to their more expressive and adeptness in handling unstructured data, making them an ideal candidate for predicting the frictional contact network.
\textcolor{black}{
Traditional methods such as graph-theoretical approaches or force/distance criteria rely on predefined rules to identify particle contacts and are unable to capture the underlying nonlinear and multi-scale interactions. By contrast, the GNN learns these relationships directly from the data, capturing hidden features and complex patterns without manual thresholds or rules. This adaptability makes GNNs highly effective for modeling complex systems with interactions that are difficult to parametrize.}
Most studies on particulate suspensions have focused only on the dilute limit~\cite{rossi2022identification, binel2023online}, not of interest here. 
%
In the dense limit, recently Mandal \textit{et al.}~\cite{Mandal_2022} employed GNN in dry granular matter, demonstrating the network prediction in both frictionless and frictional materials from the undeformed structure. 
However, the simulations were performed for relatively high volume fractions close to or above jamming, which is not feasible for rigid particles, and when tasked with an extrapolation setting, the accuracy drops sharply. 

This study addresses the previously mentioned challenges of predicting the structure of FCN in dense suspensions only with the knowledge of the relative distance between particles by employing a robust GNN approach to a well-established simulation approach for dense suspensions~\cite{Mari_2014, Mari_2015, mari_discontinuous_2015, Singh_2018, Singh_2019, Singh_2020, Singh_2022}.
By comparing the predictions from the ML technique with the simulation results, it is demonstrated that the GNN model can accurately predict the frictional contacts under applied stress.
In addition, the aforementioned capabilities hold for many values of applied stress $\sigma$, system size $N$, and particle sizes, even when the machine has not seen these conditions.
Finally, this work can help make meaningful predictions about suspension properties for computationally and experimentally challenging cases (large system sizes, bidisperse packings, etc.) while being trained on less challenging conditions within reasonable computational cost and time. 

\section{Methodology}
\begin{figure}
    \includegraphics[trim = 0mm 180mm 270mm 0mm, clip,width=0.49\textwidth,page=2]{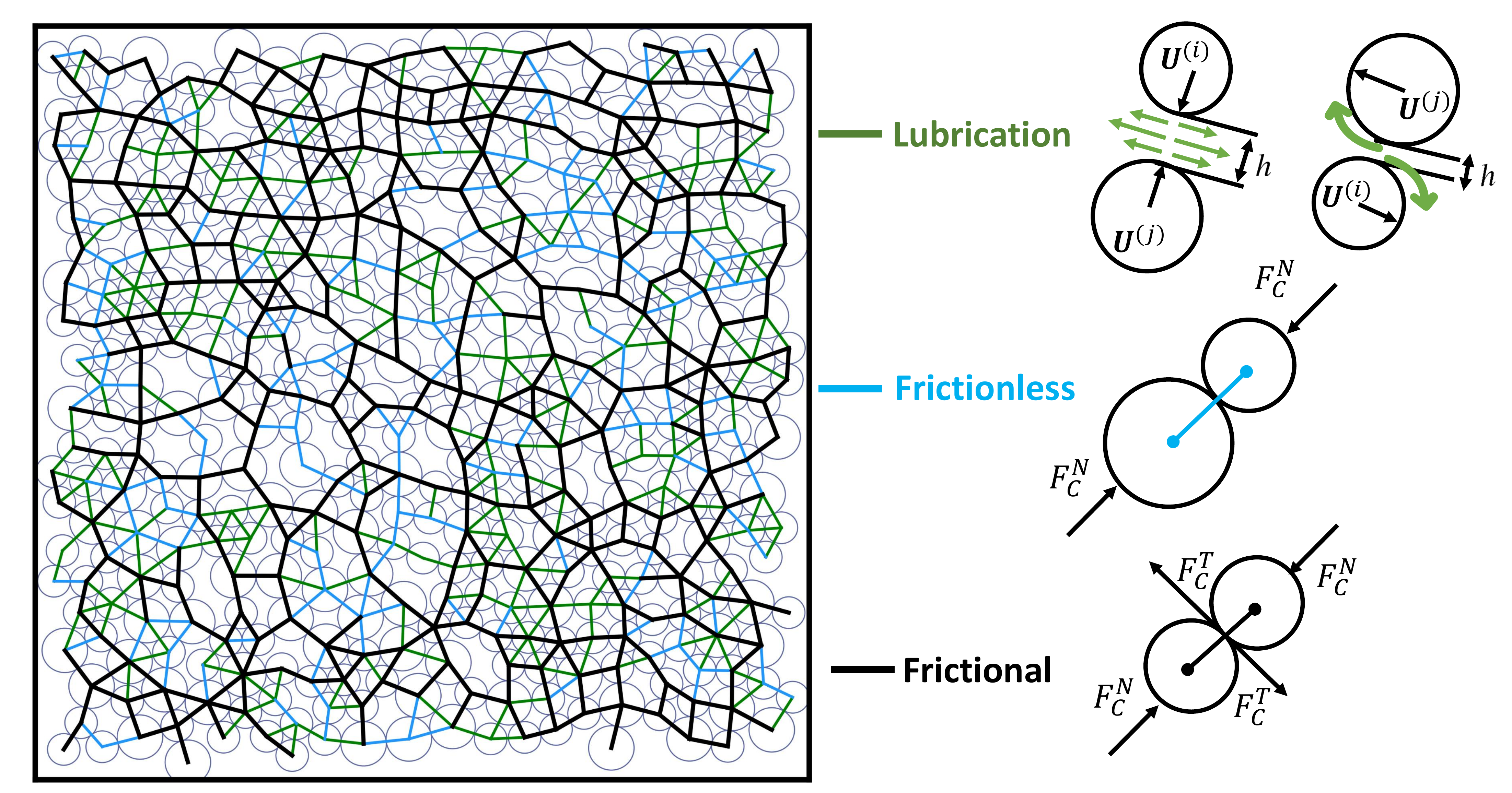}
    \caption{
    \textbf{Contact network inferring all forces.} Snapshot of contact network for $\phi=0.76$ and $\sigma/\sigma_0$ = 1. Green, blue, and black lines depict lubrication, frictionless, and frictional interactions, respectively.
    }
    \label{fig:1}
\end{figure}

\textbf{{Simulating dense suspensions}}:
Although the real-world dense suspensions are three-dimensional, related prior works have shown the flow behavior for 2D and 3D to be similar if the volume fraction $\phi$ is appropriately scaled~\cite{Nabizadeh_2022, Gameiro_2020}.
Thus, two-dimensional simulations (a monolayer of spheres) are performed for clarity, simplicity, computational efficiency, and as a first demonstration of the application of GNN to predict the FCN.
The simulation scheme (LF-DEM) integrates two modeling approaches: Lubrication flow (LF) and the discrete element DEM method from dry granular materials~\cite{Seto_2013, Mari_2014, Morris_2020, Singh_2020}. 
The particle motion is considered to be inertialess, that is, particles obey the overdamped equation of motion 
\begin{equation}
0 = \vec{F}_{\mathrm{H}}(\vec{X},\vec{U}) + \vec{F}_{\mathrm{C}}(\vec{X})~,
\end{equation}
where $\vec{X}$ and $\vec{U}$ refer to the particle position and velocities, respectively. Here, $\vec{F}_{\mathrm{H}}$, $\vec{F}_{\mathrm{C}}$ denote hydrodynamic and contact forces, respectively. The hydrodynamic force includes one body Stokes drag and two body lubrication forces. The lubrication force is regularized allowing the particles to make contact as the overlap ($\delta^\text{(i,j)} = a_i + a_j - |\vec{r_i} - \vec{r_j}|$) becomes positive~\cite{Ball_1995, Mari_2014}. Here, $a_i$ and $a_j$ refer to particle radii, and $\vec{r_i}$ and $\vec{r_j}$ represent the position vectors of the center of particles $i$ and $j$, respectively.
The contact forces include both normal and tangential frictional forces, i.e.,  $\vec{F}_{\mathrm{C}} = \vec{F}_{\mathrm{C}}^N+\vec{F}_{\mathrm{C}}^T$.
The contact force is modeled using traditional Cundall \& Strack \cite{Cundall_1979} and following the algorithm described by Luding~\cite{Luding_2008}. We employ linear springs in normal $k_n$ and tangential $k_t$ directions to model contacts between particles, which are tuned such that the maximum scaled particle overlap does not exceed 3\% and the rigid particle approximation is satisfied~\cite{Singh_2015}.
In our simulation, we do not employ the dashpot; instead, the hydrodynamic resistance provides energy dissipation~\cite{Mari_2015}. 
The tangential component of the contact force satisfies Coulomb's friction law $|\vec{F}_{\mathrm{C}}^t \le \mu |\vec{F}_{\mathrm{C}}^n|$ with $\mu=0.5$ being the static friction coefficient. 
The Critical Load Model (CLM) is used to introduce rate dependence, where the normal force ${F}_{\mathrm{C}}^N \ge F_0$ is needed to activate interparticle friction, giving a characteristic stress scale $\sigma_0 = F_0/a^2$. This implies that solely negative gap (or positive overlap$\delta$) between particles is not representative of the frictional contact between them.
%
%
A series of stress-controlled simple shear flows are simulated for $N=400$ to 5000 non-Brownian bidisperse particles with Lees-Edwards periodic boundary condition in a unit cell with particle size ratio $\Delta = R_L/R_S \in [1.4, 6]$ and different volumetric mixing ratios $\alpha = V_S/(V_L+V_S) \in [0.1,0.9]$. Here, $R_S$ ($R_L$) and $V_S$ ($V_L$) are the radii and volume of small (and large) particles, respectively.
%
This work focuses on the frictional contact network, which is the dominant contribution to viscosity at high packing fractions and is the primary driver for DST and SJ~\cite{Mari_2014, Singh_2020, Seto_2013, Ness_2016}.
Figure~\ref{fig:1} represents a typical contact network resulting from the simulations. The colors represent different types of interactions with green, blue, and red lines representing lubrication, frictionless, and frictional forces, respectively.


\begin{figure*}[hbt!]
    \includegraphics[trim = 0mm 0mm 20mm 5mm, clip,width=0.95\textwidth,page=1]{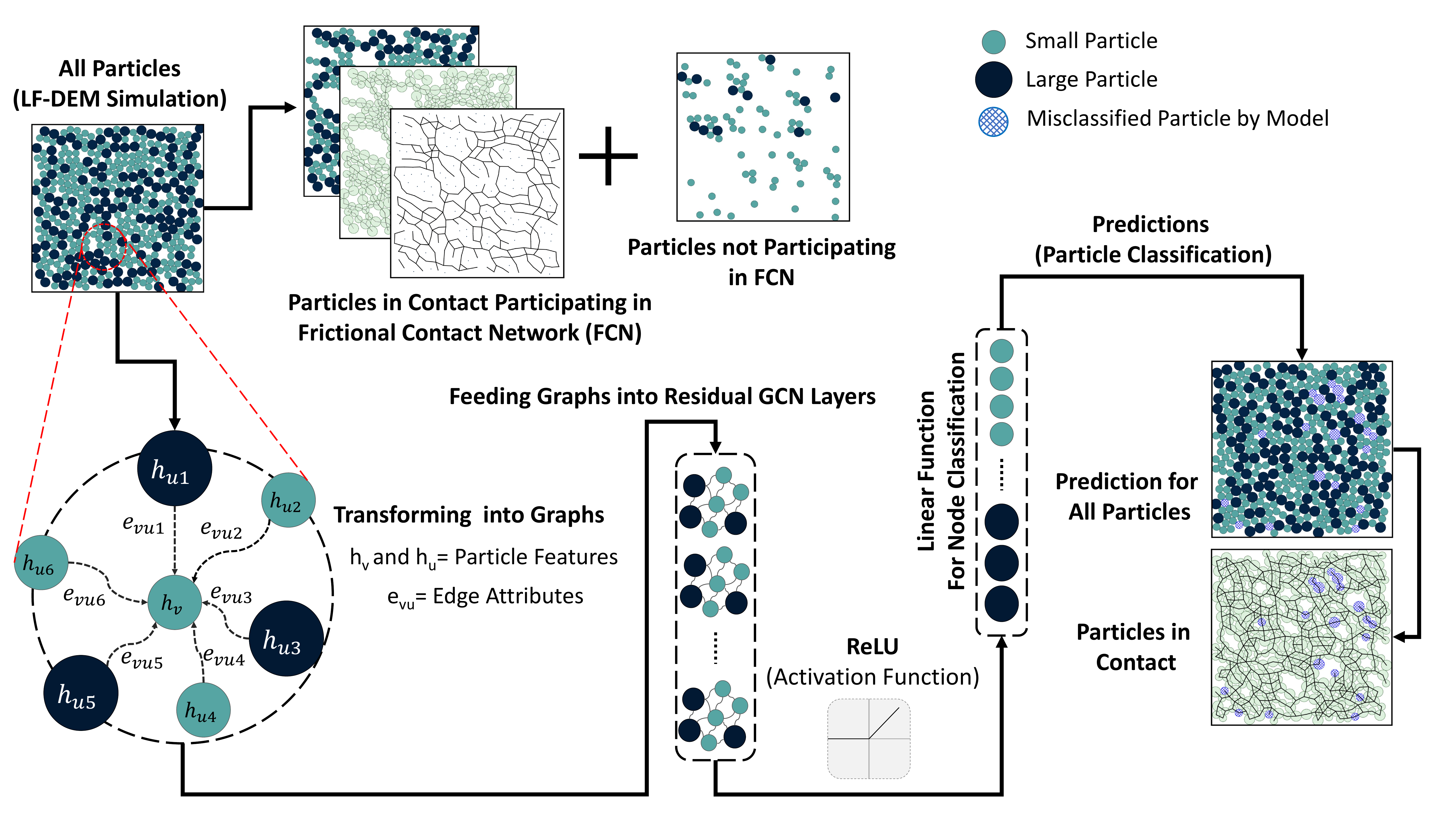}
    \caption{\textbf{Schematic detailing simulation snapshot and the Graph Neural Network (GNN) for predicting Frictional Contact Network (FCN).} Initially, necessary snapshots for ML consisting of information of all particles (whether the particles participate in the FCN or not) are driven through LF-DEM simulation technique which then is transformed into a graph comprising its constituent particles and the frictional contact network. These particles and the chain network, analogous to nodes and edges in the graph, undergo processing through residual convolutional layers. The message-passing process is done in these layers, employing an aggregation function to update information across nodes and edges. Subsequently, a non-linear activation function (ReLU) is applied to enhance the model's prediction capabilities, particularly in capturing complex relationships. A linear function is applied to assign each node to a specific class for classification. To mitigate overfitting, regularization techniques such as dropout are employed at the final stage of the model.
    }
    \label{fig:fig2}
\end{figure*}

\textbf{Machine learning methodology}: 
In this work, the concept is adapted from Li \textit{et al}., which made very deep training of graphs based on graph convolutional neural network (GCN) possible by incorporating residual and dense connections, along with the use of dilated convolutions to remedy vanishing gradient problem~\cite{Guohao_2020, li_2019deepgcns}. Their technique called deep graph convolutional neural network (DeepGCN) is used, and then the models are optimized by training them at different hyperparameters. Compared to the traditional GCN, this deep learning method provides more reliable and deeper training and an interpretable representation of large graphs and node property prediction tasks in dense particulate systems~\cite{Guohao_2020}. 
This node classification approach identifies particles in frictional contacts that participate in FCN. 

Figure~\ref{fig:fig2} illustrates the process and architecture of the DeepGCN. Initially, configurations are generated through our simulation scheme, which contains details regarding all the particles involved. These data includes information that relates to all the particles irrespective of their participation in the FCN.
%
\textcolor{black}{To train the model, first, the dynamic simulation dataset consisting of all particles is transformed into graphs by treating particles as nodes ($h_v$ and $h_u$ are node features for a specific particle and neighborhood particles, respectively) and drawing edges ($e_\text{vu}$, edge attributes) between them to represent their interactions (both frictionless and frictional).} The node features include the particle radius, and  
the edges representing interactions called edge attributes contain information about the distance between particles ($r_\text{ij}$), $x$ and $y$ components of vector $r_\text{ij}$, and the Sine and Cosine of the angle $r_\text{ij}$ makes with the x-axis (flow direction).
The DeepGCN model consists of $N_l$ layers that incorporate the residual connections. 
A node’s feature vector is updated in each residual layer through a message-passing process, aggregating information from its neighbors, including nodes and connected edges within the graph. A nonlinear activation function, ReLU, followed by a linear classification function (for node or particle classification), is applied to the output to predict the probability of each particle being a part of the FCN. 
%
The detailed equation for DeepGCN is as:

\begin{equation}
    h_v^{l} =  h_v^{(l-1)} + \sum_{u \in \mathcal{N}(v)} f(h_u^{(l-1)}, h_v^{(l-1)}, e_{uv})~.
    \label{eq:3}
\end{equation}
Here, $h_v^{(l)}$ is the updated node feature or hidden state for node $v$ at layer $l$, $ \mathcal{N}(v) $ denotes the neighborhood of node $v$, that is, the set of nodes connected to $v$, $f$ is a function that takes as input the features of neighboring nodes, i.e.,  $h_v^{(l-1)}$ and $h_u^{(l-1)}$ and their edge features $e_\text{uv}$, concatenating the features and applying a non-linear aggregation function, and $h_v^{(l-1)}$ is the original node feature that is added to the output of the GCN. More details regarding the mechanism and architecture of DeepGCN are provided in the supplemental document~\cite{NoteX}.

Figure~\ref{fig:3} illustrates the loss function and accuracy as a function of the epochs during model training. 
%
During training, the model's parameters are adjusted iteratively to align the predicted outputs more closely with the actual targets through the loss function (details about Binary Cross Entropy (BCE) are available \href{https://pytorch.org/docs/master/generated/torch.nn.BCEWithLogitsLoss.html\#torch.nn.BCEWithLogitsLoss}{here}). We employed the Adam optimizer with a learning rate of 0.005 to minimize the loss function on the training set. The accuracy is the ratio of correctly classified nodes (particles) to the total number of nodes in the evaluation set. Subsequently, we assess the network's performance on a separate and independent test sets.
%
Our training dataset comprised 320 configurations (80\% of our overall dataset), whereas a set of 80 configurations (20\% of the dataset) was reserved for testing performance of the model. 
We stop the training process when the loss value does not improve after 15 epochs by saving the last iteration parameters at which the loss is minimum. 
Subsequently, we employed 400 previously unseen configurations for validation purposes well beyond the initially seen training conditions. 
%
These predictions derived the reported average prediction accuracy of the FCN. In addition to average accuracy, we measured average precision measure, recall, F1 score, AUC, and specificity, which are presented and discussed in Table 1 in the supplemental document~\cite{NoteX}.

\begin{figure}
    \includegraphics[trim = 0mm 250mm 210mm 0mm, clip,width=0.95\textwidth,page=4]{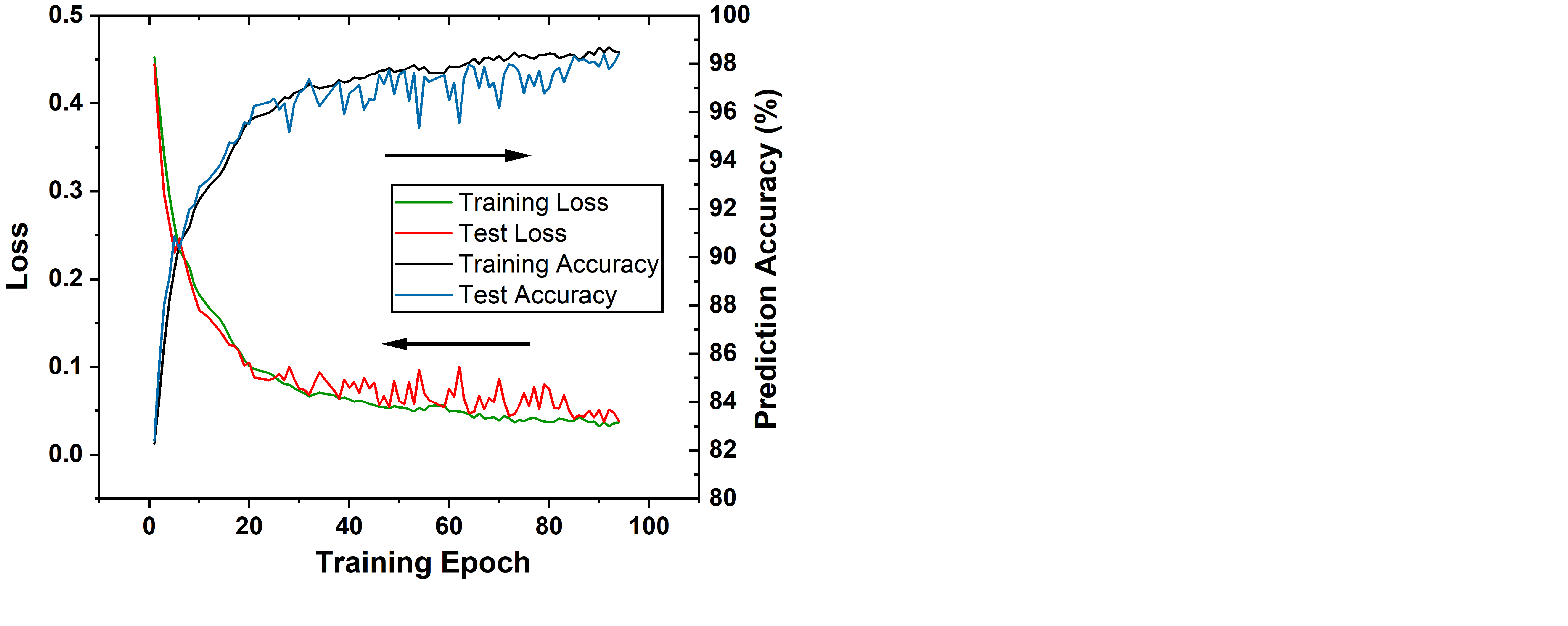}
    \caption{
        \textcolor{black}{\textbf{An example of prediction accuracy and Loss function (Binary cross entropy) for training the model.} The model is trained at $\phi$ = 0.80 and $\sigma/\sigma_0$ = 10 with 320 graphs and a set of 80 graphs for testing using the early stop technique that prevents overfitting.}
    }
    \label{fig:3}
\end{figure}

\begin{figure}
    \includegraphics[trim = 0mm 170mm 360mm 0mm, clip,width=0.55\textwidth,page=3]{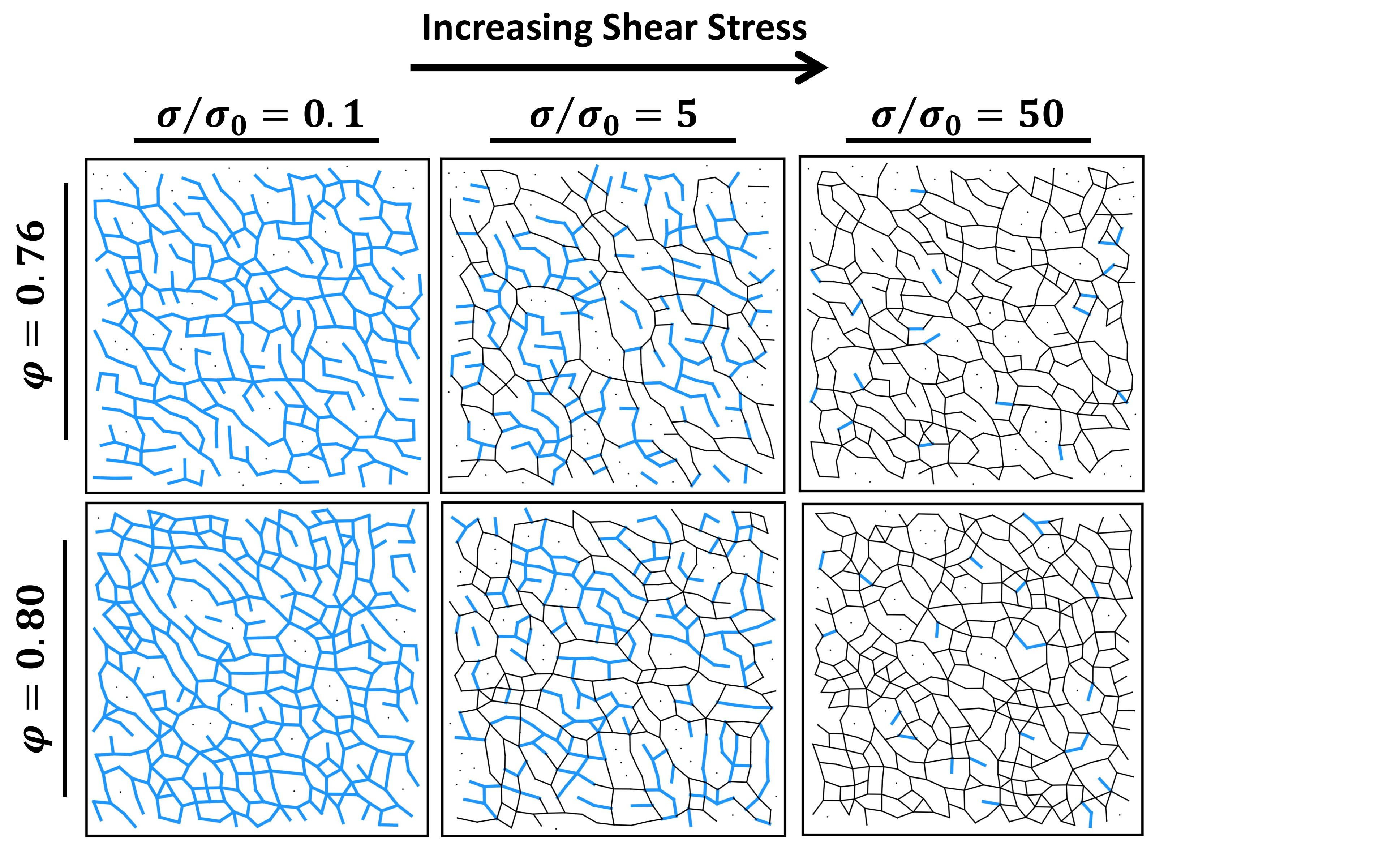}
    \caption{ \textcolor{black}{\textbf{Development of frictional contacts with increasing shear stress.} Snapshots of the contact network for both frictionless (normal force below the critical threshold $F_0$) shown as blue lines and frictional contacts (normal force above the critical threshold $F_0$) as black lines. Snapshots corresponding to simulations with $\sigma/\sigma_0 =$ 0.1, 5, 10, and 50 (from left to right) are presented for $\phi=0.76$ (top) and $\phi=0.8$ (bottom). }}
    \label{fig:4}
\end{figure}

\begin{figure*}
    \includegraphics[trim = 0mm 5mm 280mm 0mm, clip,width=0.85\textwidth,page=5]{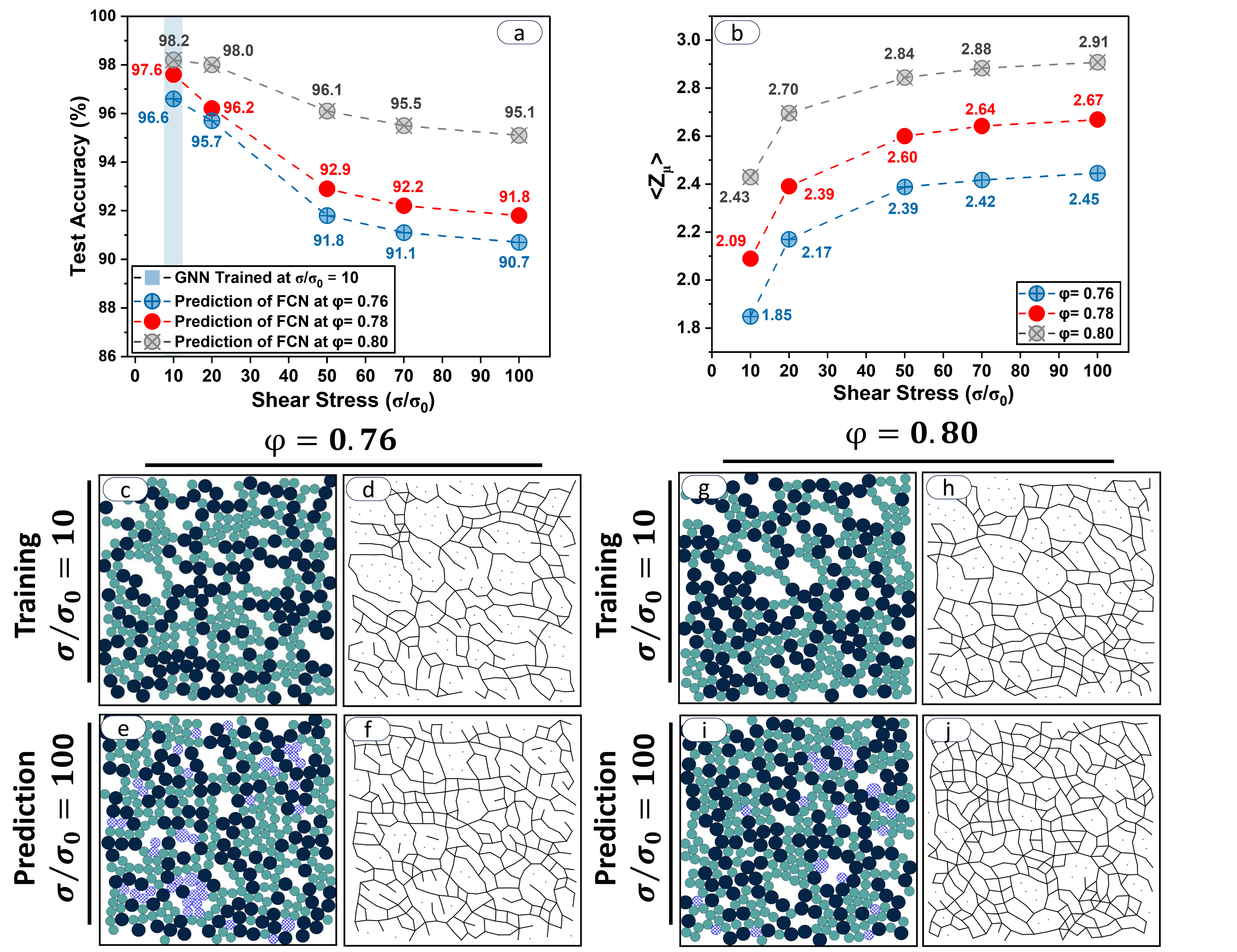}
    \caption{\textbf{Robustness in the prediction of FCN for a prototypical shear thickening suspension.} The GNN models are trained separately on the simulation data set at a fixed stress $\sigma/\sigma_0=10$ and at different packing fractions $\phi$ for $N=400$ and bidispersity $(\Delta,\alpha = 1.4, 0.5)$. 
    (a) Prediction accuracy for FCN for different values of $\phi,\sigma/\sigma_0$, with the GNN conditioned at $\sigma/\sigma_0=10$ for each $\phi$.
    (b) Frictional coordination number $\langle Z_\mu \rangle$ as a function of stress for various values of $\phi$. For the sake of simplicity, here, only the visualization consisting of particles participating in FCN are provided where
    (c-d) are visual of the training condition with $\phi=0.76$, showing (c) sheared packing at $\sigma/\sigma_0=10$ with particles participating in the FCN along with 
 (d) the black lines showing the frictional network. (e) and (f) Predicted configuration at $\sigma/\sigma_0=100$ with (e) particles participating in the FCN along with the misclassified particles by the model shown in lattice-patterned blue and (f) the predicted FCN. Misidentified particles are shown in \textcolor{black}{lattice-patterned blue}. 
    (g-h) Same as (c-d) but for $\phi=0.8$ and $\sigma/\sigma_0=10$.
    (i-j) the corresponding network based on (g-h).}
    \label{fig:fig5}
\end{figure*}

\begin{figure}
    \includegraphics[trim = 0mm 220mm 170mm 0mm, clip,width=0.90\textwidth,page=6]{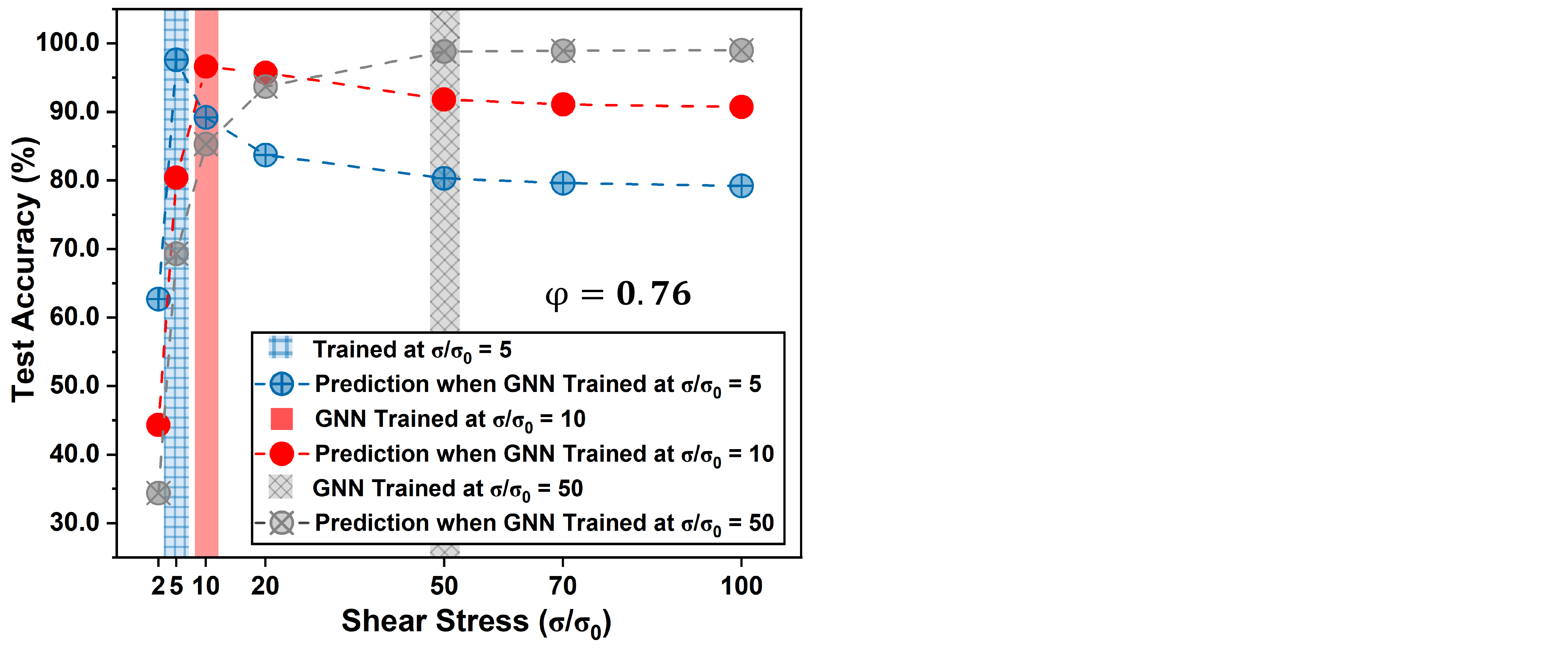}
    \caption{
    \textbf{Comparison of prediction accuracy of the GNN when trained at different stresses} Prediction accuracy as a function of $\sigma/\sigma_0$ while the models have been trained separately at $\phi = 0.76$ and different fixed $\sigma/\sigma_0$ = 5, 10, and 50.}
    \label{fig:fig6}
\end{figure}

\section{Results}
\textbf{Evolution of Frictional Contact Network under Shear Stress}: 
Before turning to the results on the accuracy of the model, Fig.~\ref{fig:4} shows the evolution of the contact network obtained from the simulations at packing fractions of $\phi=0.76$ and 0.80 with increasing stress $\sigma/\sigma_0$ from left to right. Line segments connect the centers of two contacting particles and are color-coded according to the type of force they experience; we show frictionless contacts (blue line) and frictional (black line). At the lowest stress considered here $\sigma/\sigma_0=0.1$, only frictionless contacts (blue bonds) are observed. At higher stresses $\sigma/\sigma_0=5,10$, both frictionless (blue) and frictional (black) contacts are observed. Eventually, at the highest stress  $\sigma/\sigma_0=100$, the suspension is fully frictional (only black lines are present). This visual observation is consistent with the literature~\cite{Mari_2014} and our results on coordination number (Fig. S1). Note that for a given, constant packing fraction $\phi$, frictional coordination number $Z_\mu$ in a fully frictional state is larger than the frictionless contacts at low stress, non-thickened state. This suggests that with an increase in stress, as particle contacts become frictional, the suspension rearranges into a distinct microstructure. Hence, it is non-trivial to predict the FCN even knowing particle overlap without the critical load force $F_0$.

The following describes two striking features of the DeepGCN scheme for predicting frictional contacts: robustness and scalability.

\textbf{Robustness of DeepGCN Model}: First, the DeepGCN method is shown to be highly robust in predicting the frictional contacts in a dynamical system, where the network continuously forms and breaks due to the bulk shearing motion (Fig.~\ref{fig:fig5}).
$N=400$ particles are used at packing fractions $\phi=0.76, 0.78, 0.8$ with $\Delta=1.4$ and $\alpha=0.5$.
The full rheological flowcurves ($\eta_r(\phi,\sigma/\sigma_0)$) and the frictional coordination number ($Z_{\mu}(\phi,\sigma/\sigma_0)$) are presented in the Supplemental Material (Fig. S1)~\cite{NoteX}. In short, the suspension undergoes CST for  $\phi=0.76, 0.78$ and DST into a nearly jammed state for $\phi=0.8$.
\textcolor{black}{The presented two-dimensional rheological flowcurves are also compared with the literature results in SI. We show qualitative agreement with previous three-dimensional simulations (Fig. S4) and experimental data on the silica colloids (Fig. S5), once the packing fraction is properly scaled with respect to $\phi_J^\mu$.}

The model is trained at $\sigma/\sigma_0=10$ for each $\phi$, and predictions are made for other applied stresses. 
%
Remarkably, despite the sensitivity of the FCN to both the stress and packing fractions, the model is capable of accurate predictions ($>90\%$) for all values of ($\sigma/\sigma_0$, $\phi$) (Fig.\ref{fig:fig5}~a).
In Figs.~\ref{fig:fig5}~e-f and i-j, the predictions with numerical simulation results at $\sigma/\sigma_0=100$ are visualized where the models are trained at $\sigma/\sigma_0=10$ (Figs.~\ref{fig:fig5}~c-d and g-h) for the corresponding $\phi=0.76$ and 0.8, respectively. 
%
For simplicity, only the particles that have at least one frictional contact (Figs. ~\ref{fig:fig5}c, e, g, and i) together with their FCN (Figs. ~\ref{fig:fig5}d, f, h, and j) are shown, and the configurations do not depict all particles ($N=400$). 
The predictions depict ground truth for direct simulations, where the misclassified particles by the model are highlighted in red. Besides that, larger (smaller) particles are represented in darker (lighter) colors.
As can be seen, only a few particles evade the predictions of the model. Especially for $\phi=0.8$, the training stress ($\sigma/\sigma_0=10$) is in an unjammed flowing state, while the prediction stress ($\sigma/\sigma_0=100$) is nearly in the shear-jammed state, yet an accuracy of 95\% is achieved, highlighting the strength of the model. 
%
{This accuracy at $\phi = 0.80$ is expected since the model has ``seen" the complex structure; hence, it can accurately predict the complex features.} 
Figure~\ref{fig:fig5}b shows $Z_\mu$ for the values of $\sigma/\sigma_0$ of interest here (results for all stress values in~\cite{NoteX} (Figs. S1)).
With increasing stress, the FCN becomes increasingly interconnected, spanning both compressive and tensile directions (Ref. Fig.~\ref{fig:3}). The decrease in prediction accuracy at higher stresses suggests the difficulty of predicting FCN, highlighting the increase in complexity at higher stresses. 
The results of these rigorous checks demonstrate that the model maintains a highly accurate performance, with extrapolated predictions for unseen conditions, without prior knowledge of interparticle forces, and with only information on relative distance solely at a fixed stress value. 

A natural question arises regarding the choice of $\sigma/\sigma_0 = 10$ to train the model on. Figure~\ref{fig:fig6} illustrates the test accuracy results when the model is trained at different values of $\sigma/\sigma_0$ for a fixed $\phi = 0.76$. Although the predictions of the model trained at $\sigma/\sigma_0$ = 5 are unsatisfactory, the predictions improve significantly when the model is trained at $\sigma/\sigma_0$ = 10 and 50. These results align with the fact that the number of frictional contacts increases with increasing stress. Hence, at higher stresses, it is easier for the model to predict whether a given particle is in frictional contact.
It is important to note that this is an extrapolation task, which is inherently more challenging for the model, particularly given that the rheological properties of suspensions vary with $\sigma/\sigma_0$ (Fig. S1b).
%
%
To explore the physical rationale behind this observation, Fig. S3 illustrates the evolution of the contact network and corresponding structure factor with increasing stress. As shown already in Figs.~\ref{fig:3} and S1a, the number (or fraction) of frictional contacts begins to saturate (frictionless contacts diminish) in the limit of $\sigma/\sigma_0 \ge 50$, which implies that the physics is fully dominated by frictional contacts in that regime. 
Although the evolution trend appears similar, a larger number of contacts (denser FCN) are observed for a higher packing fraction $\phi=0.8$.

%

\begin{figure*}[hbt!] 
    \includegraphics[trim = 0mm 290mm 200mm 0mm, clip,width=0.95\textwidth,page=7]{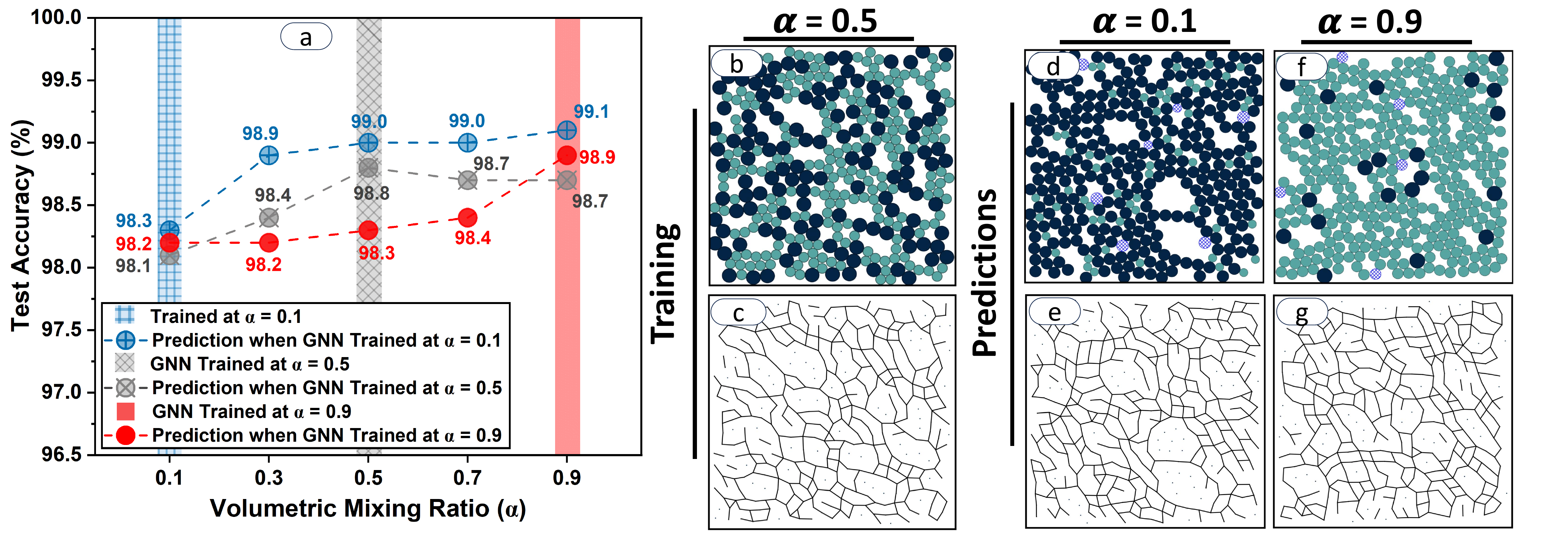}
    \caption{\textbf{Robustness in FCN predictions for various volumetric mixing ratios $\alpha$}. All the configurations are generated for $\phi=0.76$, $\Delta=1.4$, and sheared at fixed stress of $\sigma/\sigma_0=50$. The GNN model is trained separately with the configurations consisting of 400 particles at different $\alpha=$ 0.1, 0.5, and 0.9 and can also robustly predict the FCN at other values of $\alpha$. \textcolor{black}{(a) Test accuracy results, }
     (b-c) The configuration on which the training is conditioned with (b) showing sheared packing with particles participating in the FCN where the dark (and light) blue depicts larger (and smaller) particles, and (c) frictional network. 
     (d-e) The predictions for $\alpha=0.1$: (d) showing the configuration with particles in contact and the misclassified particles in red, (e) black lines showing the corresponding FCN.
     (f-g) Same as (d-e) but for the configuration with $\alpha=0.9$. Note that (b), (d), and (f) only show particles with at least one frictional contact.
    }
    \label{fig:fig7}
\end{figure*}

Experimental and simulation studies have shown that bidispersity significantly affects the rheology and microstructure of dense particulate systems and colloidal gels~\cite{Maranzano_2001, Pednekar_2018, petit2020additional, Guy_2020, Singh_2024, Waheibi_2024}. \textcolor{black}{The dependence of viscosity ($\eta_r (\alpha)$) and frictional coordination number ($Z_\mu (\alpha)$) on $\alpha$ are presented in the Supplemental Material (Fig. S2)~\cite{NoteX}.}
%
Hence, it is tempting to ask whether the presented GNN model can predict the FCN for different values of $\alpha$ for a constant $\Delta$.
Figure~\ref{fig:fig7}a demonstrates the robustness of the GNN model in predicting the FCN at various values of $\alpha$ for a fixed size ratio $\Delta = 1.4$ with an accuracy exceeding 98\%.
Figure~\ref{fig:fig7}b shows an example of the training set consisting of particles in frictional contact, and Fig.~\ref{fig:fig7}c shows the corresponding FCN.
This visualization shows only particles with at least one frictional contact.
Figures~\ref{fig:fig7}, d \& e show examples for visualization of the model predictions (only particles in contact and the misidentified particles depicted in red) for $\alpha=0.1$ \& 0.9 (on previously unseen configurations) with remarkable accuracies along with their FCN (Fig.~\ref{fig:fig7} e,g). 
Visualizations for all particles, the predicted structure factor of the exact frictional contacts, and the absolute error of the prediction and direct simulation results are analyzed and shown in Fig. S4. 
This consistent accuracy is achieved (Fig.~\ref{fig:fig7} a) for all values of $\alpha$ even though the contact network has a distinct structure (dominated by small or large particles for the two values of $\alpha$).
The accuracy is \textit{robust} to the choice of the training value of $\alpha$ ((0.1, 0.5, or 0.9)).
%
%
This implies that the presented GNN model can predict the FCN for a distinct system despite having information from only a limited number of particles, mostly of one type. 

\begin{figure*}[hbt!] 
    \includegraphics[trim = 0mm 270mm 80mm 0mm, clip,width=0.99\textwidth,page=8]{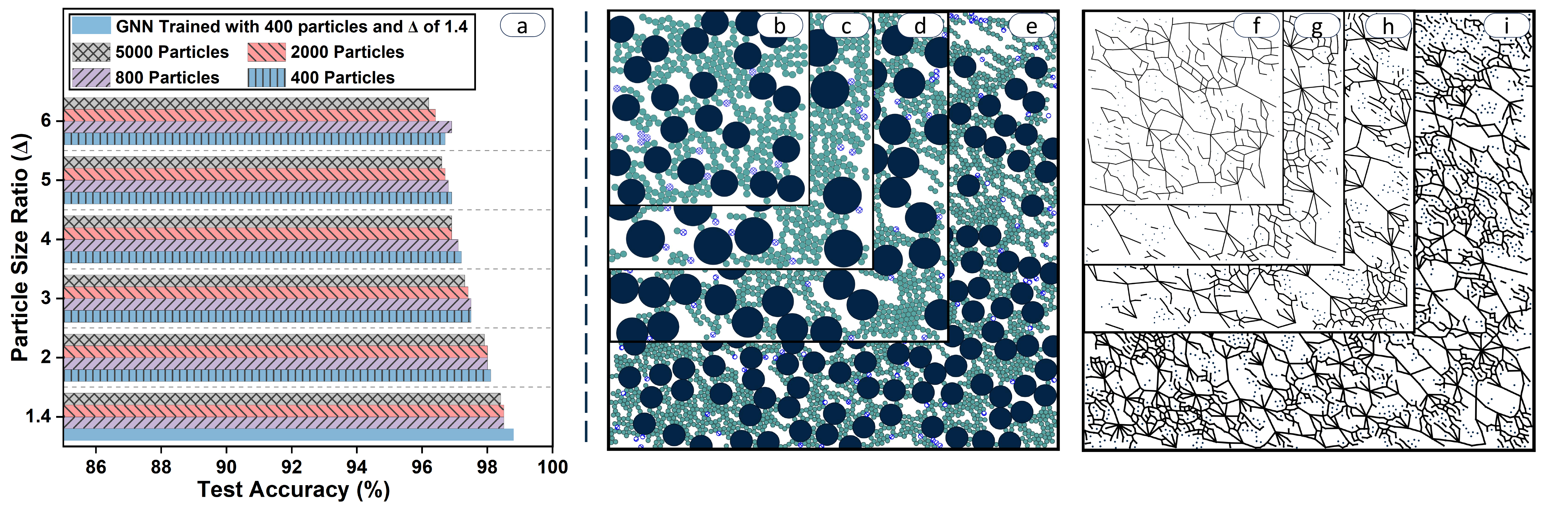}
    \caption{\textbf{Scalability in predicting FCN across different system sizes and varying particle size ratios $\Delta$.} All configurations sheared at a constant scaled stress $\sigma/\sigma_0 =50$, and packing fraction $\phi=0.76$ and volumetric mixing ratio $\alpha=0.5$. The model is trained on configurations with $N=400$ for $\Delta=1.4$ (Figs.~\ref{fig:fig7}b and c).
     (b-e) The prediction of particles participating in FCN along with misclassified particles highlighted in \textcolor{black}{lattice-patterned blue} at different systems, with (b) $\{N,\Delta\} = \{400, 4\}$, (c) $\{N,\Delta\} = \{800, 6\}$, (d) $\{N,\Delta\} = \{2000, 6\}$, and (e) $\{N,\Delta\} = \{5000, 5\}$. (f-i) Displaying the corresponding frictional contact network for b-e.
    }
    \label{fig:fig8}
\end{figure*}

\textbf{Scalability of DeepGCN Model}:
%
Next, the scalability of the ML model is demonstrated to make predictions for larger system sizes and complexities, such as different particle size ratios, \textcolor{black}{despite the distinct rheology and microstructure}.
\textcolor{black}{The relative viscosity $\eta_r$ and frictional coordination number $Z_\mu$ as a function of $\Delta$ for suspensions with different particle numbers are shown in Fig.~S3~\cite{NoteX}}.
This ability of the model is tested by training it on a rather small system size ($N=400$) at $\sigma/\sigma_0=50$ and $\phi=0.76$ (Figs.~\ref{fig:fig7}b and c) and then testing the predictions with not only a larger $N$ but also a larger size ratio $\Delta$. 
Figure~\ref{fig:fig8} demonstrates the scalability, showing no substantial decrease in accuracy even at 10 times the system size across very different particle size ratios (from 1.4 to 6). 
Figures~\ref{fig:fig8}b to e depict examples of the predictions of particles participating in FCN for different systems, with $\{N,\Delta\} = \{400, 4\}$, $\{N,\Delta\} = \{800, 6\}$, $\{N,\Delta\} = \{2000, 6\}$, and $\{N,\Delta\} = \{5000, 5\}$, respectively. (f-i) Displaying the corresponding frictional chain network for b-e. Here, darker particles represent particles with larger radii, and misclassified particles are highlighted in lattice-patterned blue to show the difference between predictions and ground truth.
Interestingly, larger particles are less prone to errors as they can make more contacts than smaller particles, given a larger surface area. The results show that the model can capture the relationship between particles and adapt it to large-scale systems, given only local neighborhood information. 

\textcolor{black}{
Before we close, it is important to discuss the role of dimensionality and possible extension to experimental systems. For the sake of simplicity and demonstration of application of GNN method to frictional contact network (FCN) in dense suspensions, we simulated a two-dimensional monolayer of spheres. Our results in predicting unseen FCN by GNN are promising and point towards extending the formalism to three-dimensional systems, given that the only information needed are particle positions, radii of particles and interparticle gaps. 
Nonetheless, a previous study by Gameiro et al.~\cite{Gameiro_2020} had demonstrated that the correlation between frictional network topological features and the bulk rheology are very similar in two-and three-dimensional systems. 
We also show that the two-dimensional results presented here can qualitatively reproduce the literature results based on three-dimensional simulations and experiments (SI).
%
This similarity is primarily because, in the shear thickening suspensions, the structural and kinematic features appear in the velocity and velocity gradient directions with minimal variation across the vorticity direction. 
Thus, the structural interrogations concerning the velocity and gradient directions are sufficient to describe the main structural/rheological features of the system.
This assumption might not hold in some cases in real-world flow, as an example, when orthogonal flows are superposed in the vorticity direction~\cite{Sehgal_2019}.
}

\section{Concluding remarks}
This study demonstrated the application of DeepGNN in predicting the frictional contact network (FCN) structure in dense suspensions by utilizing fixed configurations under simple shear. 
The obtained DeepGNN model exhibits remarkable generalization capabilities, accurately predicting FCNs across a wide range of unseen configurations, even under conditions significantly different from the original training environment.
This highlights its \textit{scalability} and \textit{robustness} in predicting the FCN across varying system sizes $N$, bidispersity ($\Delta$ and $\alpha$), and applied stress $\sigma$.
%
This DeepGNN model offers valuable insights into the particle information and frictional contact structure of the \textcolor{black}{simulated} suspensions, without the need for explicit knowledge of interparticle interactions.
The presented GNN model is particularly appealing because of the computational expense of tracking system dynamics with many-body interactions and boundary conditions. 
Its scalability and versatility suggest its potential for predicting various physical and rheological properties in more complex real-world flows of experimental relevance.
Although the frictional force chain concept originated in a two-dimensional granular system~\cite{Majmudar_2005}, recent experimental efforts have extended to not only three-dimensional granular systems but also to dense suspensions~\cite{Pradeep_2020, Pradeep_2021} as well.

%
%
 %
 \textcolor{black}{
 Although the current study focused only on {lubrication} and frictional interactions, in principle, the model can be extended to incorporate { van der Waals, depletion, and electrostatic forces, as well as more complete hydrodynamic interactions}. Such models have potential applications in diverse particulate systems such as colloidal gels, polymer composites, foams, and granular materials  where network physics is critical towards understanding the bulk response~\cite{smith2024topological, gupta2024next, whitaker2019colloidal, bindgen2020connecting, Nabizadeh_2024}.
%
%
%
This capability will enable the exploration of large-scale, real-life systems in natural and industrial contexts with reduced computational costs and resources.}
%

\section*{Author Contributions}
\textbf{Armin Aminimajd}: Conceptualization (equal); Data curation (lead); Formal analysis (lead); Writing – original draft (lead).\\
\textbf{Joao Maia}: Conceptualization (lead); Data curation (equal); Formal analysis (equal); Writing – original draft (equal).\\
\textbf{Abhinendra Singh}: Conceptualization (lead); Data curation (equal); Formal analysis (equal); Writing – original draft (lead).

\section*{Conflict of Interest}
The authors have no conflicts to disclose.

\section*{Acknowledgements}:
%
This work used the High-Performance Computing Resource in the Core Facility for Advanced Research Computing at Case Western Reserve University.
A. S. acknowledges the Case Western Reserve University for start-up funding.
S. Rogers, M. Ramaswamy, V. Sharma, S. Root, and S. Pradeep are acknowledged for their fruitful discussions.

\vspace{2mm}

\bibliography{dst}
\bibliographystyle{apsrev4-1}

\clearpage
\begin{widetext}
\begin{appendix}
    \section*{Supplementary Information to: \\
Scalability of Graph Neural Network in Accurate Prediction of Frictional Contact Network in Suspensions}
   In this document, we provide details about (i) training method, (ii) training parameters \& process, (iii) rheological results, and (iv) further calculations complementing the ones presented in the main text. 

\section{Rheological Results}

\paragraph{Rheological results for nearly monodisperse case $\Delta=1.4$}
\renewcommand{\thefigure}{S\arabic{figure}} 

Figure~\ref{fig:s1} depicts the rheological results obtained from simulations for $\{\Delta,\alpha\} = \{1.4, 0.5\}$. This choice of $\{\Delta,\alpha\}$ ensures that we do not observe crystallization.  As shown, that increase in stress level $\sigma/\sigma_0$ leads to an increase in viscosity $\eta_r$ (Fig.~\ref{fig:s1}b), which is associated with stress-activated transition from frictionless to frictional contacts (Fig.~\ref{fig:s1}a). As the frictional contacts (solid symbols) increase with stress, the frictionless coordination number (open symbols) decreases above the onset stress of $\sigma/\sigma_0=1$. Both, the bulk response, i.e., viscosity $\eta_r$ and frictional coordination number $Z_\mu$ mirror each other. 
Figure~\ref{fig:s1}b displays rheological response, where continuous shear thickening is observed for $\phi =$ 0.76 \& 0.78, while suspension undergoes discontinuous shear thickening (DST) for $\phi=0.8$, note the S-shaped $\eta_r(\dot{\gamma})$ flow curve for $\phi=0.8$ (inset of Fig.~\ref{fig:s1})
\begin{figure*}[h]
    \includegraphics[trim = 0mm 280mm 270mm 0mm, clip,width=0.85\textwidth,page=10]{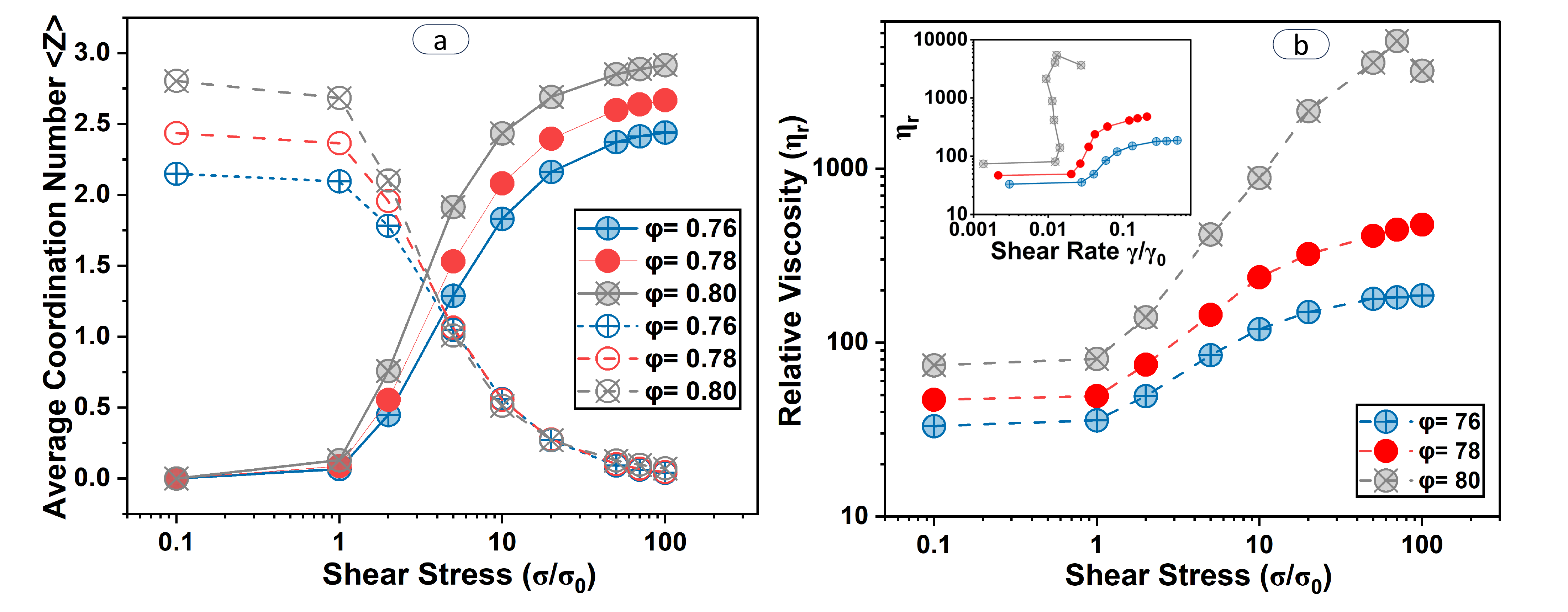}
    \caption{Evolution of microscale structures and the rheological response of the material. 
 (a) Mean number of particle-particle contacts $Z$; with solid symbols representing frictional contacts (friction is activated), open symbols representing frictionless contacts (friction is not activated).    
 (b) shear viscosity $\eta_r$ plotted as a function of shear stress $\sigma/\sigma_0$, with inset showing shear viscosity $\eta_r$ as a function of shear rate $\dot{\gamma}/\dot{\gamma}_0$. Different symbols represent different packing fractions.}
    \label{fig:s1}
\end{figure*}


\clearpage

\textcolor{black}{
\paragraph{Effect of Volumetric Mixing Ratio ($\alpha$):}
Figure~\ref{fig:s2} shows the rheological results obtained from simulations as a function of $\alpha$. We show that both $\eta_r$ and $Z_\mu$ depend on the volumetric mixing ratio.}

\begin{figure*}[h]
    \includegraphics[trim = 0mm 260mm 260mm 0mm, clip,width=0.9\textwidth,page=17]{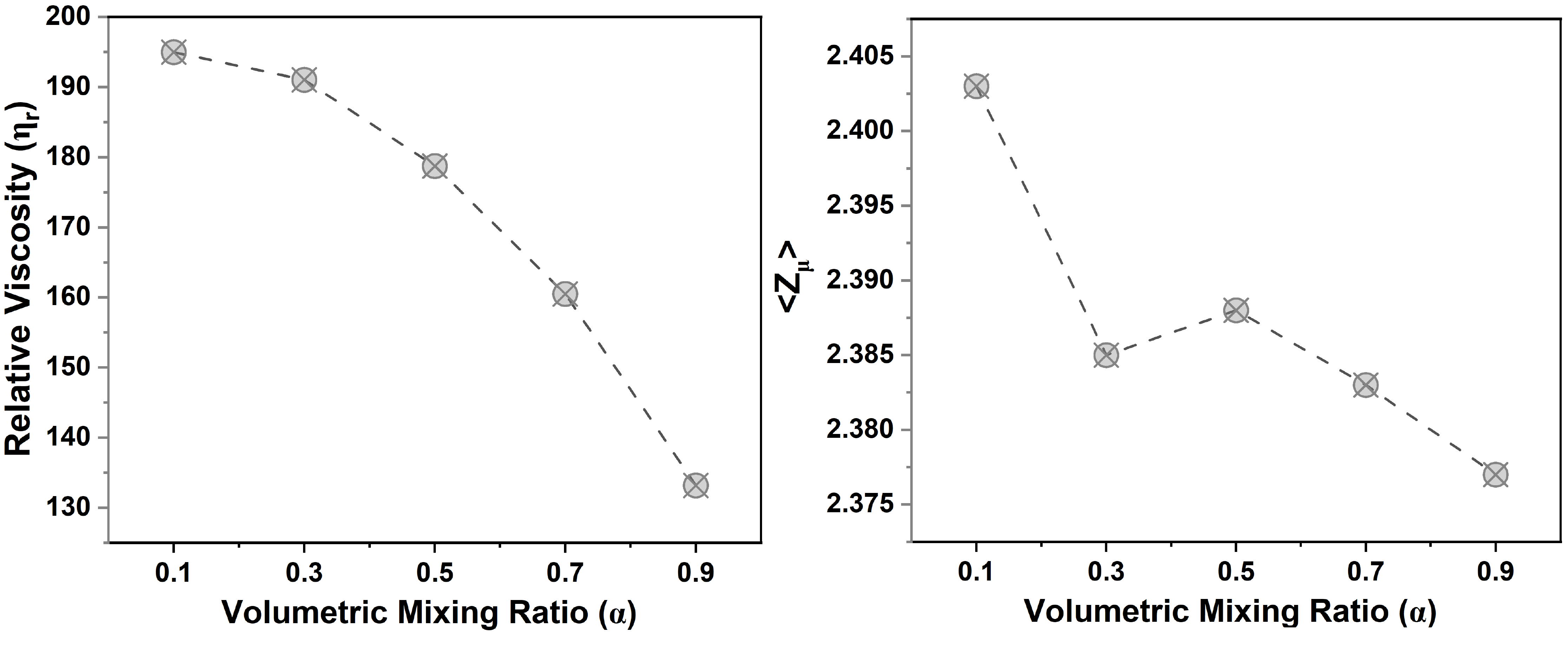}
    \caption{\textcolor{black}{Rheological and microstructural behavior of simulated suspension at various volumetric mixing ratios ($\alpha$) with left showing shear viscosity $\eta_r$ and the right plot representing the mean number of particle-particle contacts $Z$ as a function of $\alpha$ for a system at $\sigma/\sigma_0$ = 50 and $\phi$ = 0.76.}}
    \label{fig:s2}
\end{figure*}

\textcolor{black}{
\paragraph{Effect of bidispersity and system size:}
Figure~\ref{fig:s3} shows the rheological results obtained from simulations as a function of $\Delta$ for different system sizes $N$. The presented data is simulated for a fixed $\alpha$ = 0.5, $\sigma/\sigma_0$ = 50 at packing fraction $\phi=0.76$. We observe both average viscosity and frictional coordination number to decrease as a function of $\Delta$, consistent with recent 3-dimensional bidisperse simulation results~\cite{Singh_2024}. The dependence on system size $N$ only becomes noticeable for large values of $\Delta$.}

\begin{figure}[h]
    \includegraphics[trim = 0mm 250mm 200mm 0mm, clip,width=0.9\textwidth,page=18]{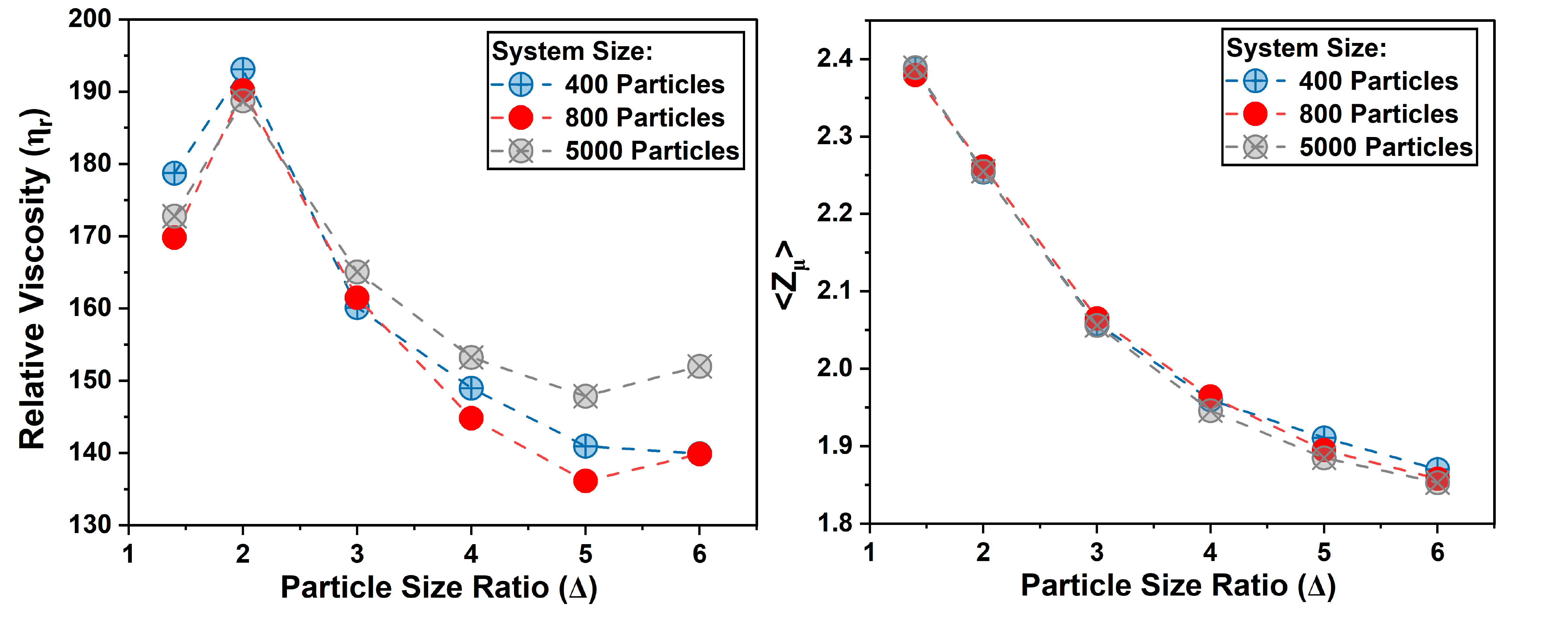}
    \caption{\textcolor{black}{Microscale structures and the rheological response of the simulated suspension at various particle size ratios ($\Delta$) and system sizes $N$. The left figure depicts shear viscosity $\eta_r$ and the right figure represents the mean number of particle-particle contacts $Z_\mu$, plotted as a function of $\Delta$ for a system with $\alpha$ = 0.5, $\sigma/\sigma_0$ = 50 and $\phi$ = 0.76. Different symbols represent different system sizes.}}
    \label{fig:s3}
\end{figure}

\clearpage
\section*{{Comparison with Literature Observations}}

\paragraph{\textcolor{black}{Comparison with Three-dimensional Simulations:}}

\textcolor{black}{
In Fig.~\ref{fig:s4}, we present the comparison between the numerical simulations for the present two dimensional system and previous three dimensional simulations presented in ~\cite{Singh_2018}. We plot $\eta_r$ as a function of scaled stress $\sigma/\sigma_0$ for both the cases for various packing fractions, where lines and symbols showing three-dimensional and two-dimensional simulations, respectively. As we show that the present two-dimensional simulations of a monolayer of spheres can quantitatively reproduce the three-dimensional simulation data as long as the packing fraction relative to $\phi_J^\mu$ is scaled correctly.}
   
\begin{figure}[h]
    \includegraphics[trim = 0mm 220mm 230mm 0mm, clip,width=0.8\textwidth,page=15]{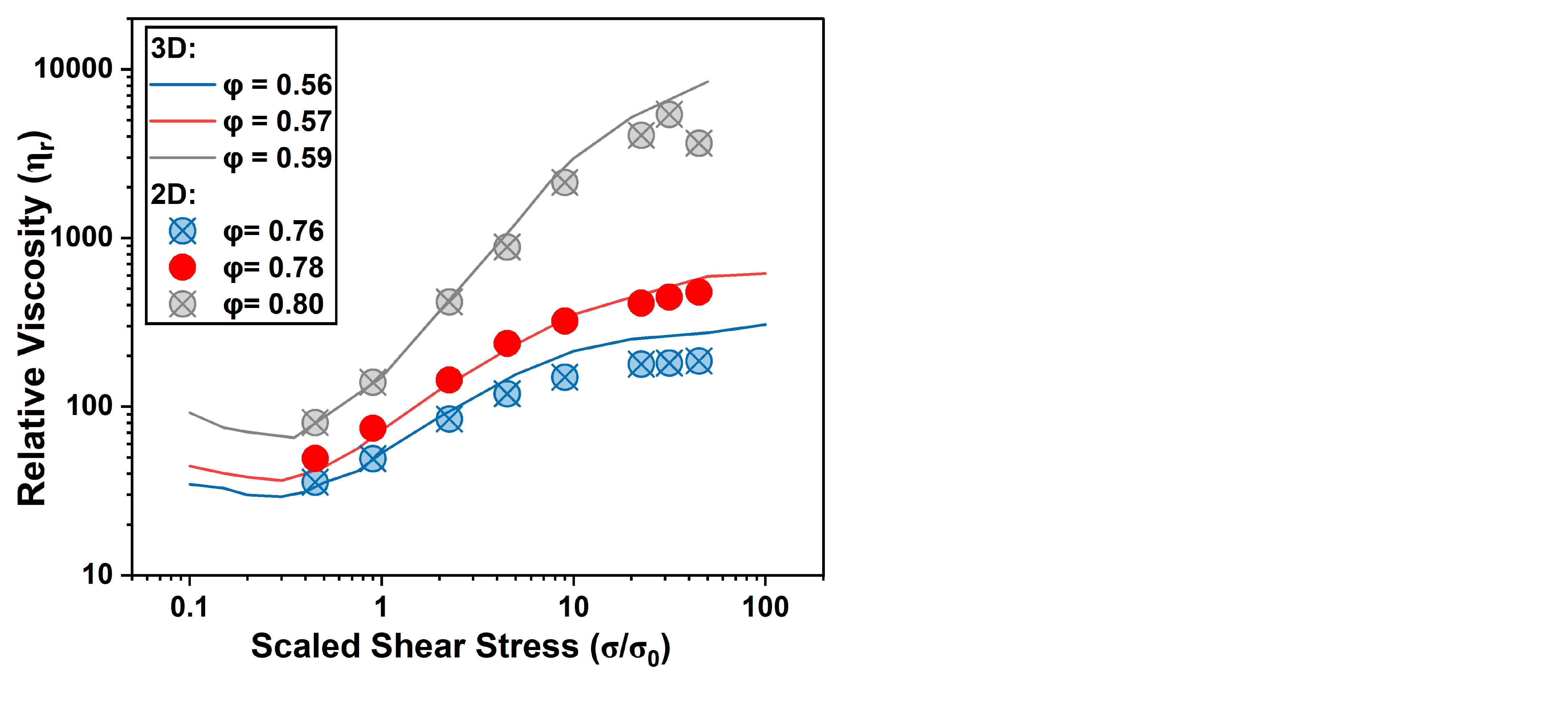}
    \caption{\textcolor{black}{Comparison of rheological behavior of simulated suspensions in two-and-three dimensions. Symbols represent the two-dimensional data from this study, while lines represent data from previous three-dimensional simulation study by Singh et al.~\cite{Singh_2018}. For the two cases, interparticle friction $\mu=0.5$ was used. 
}}
    \label{fig:s4}
\end{figure}

\paragraph{\textcolor{black}{Comparison with Experiments:}}
\textcolor{black}{
In order to quantitatively compare our simulation data sets with experiments, it is natural to ensure that the packing fraction is correctly scaled. Figure~\ref{fig:s5} shows a comparison between simulation results compared against the experimental data from Royer et al.~\cite{Royer_2016} for 1.54 $\mu$m sized silica particles (also non-Brownian). For both cases, we present the shear-thickened state viscosity, i.e., $\eta_r(\sigma/\sigma \to \infty)$. For the experimental data set, we use $\phi_J^{\mathrm{exp}}=0.592$ (as reported by Royer et al.) and in our 2D simulations we find $\phi_J^{\mathrm{sim}}=0.815$. An excellent collapse between the simulation and experimental data sets suggest that our 2-dimensional simulations represent the appropriate physics in the dense limit.}

\begin{figure}[h]
    \centering
    \includegraphics[trim = 0mm 220mm 230mm 0mm, clip,width=0.8\textwidth,page=16]{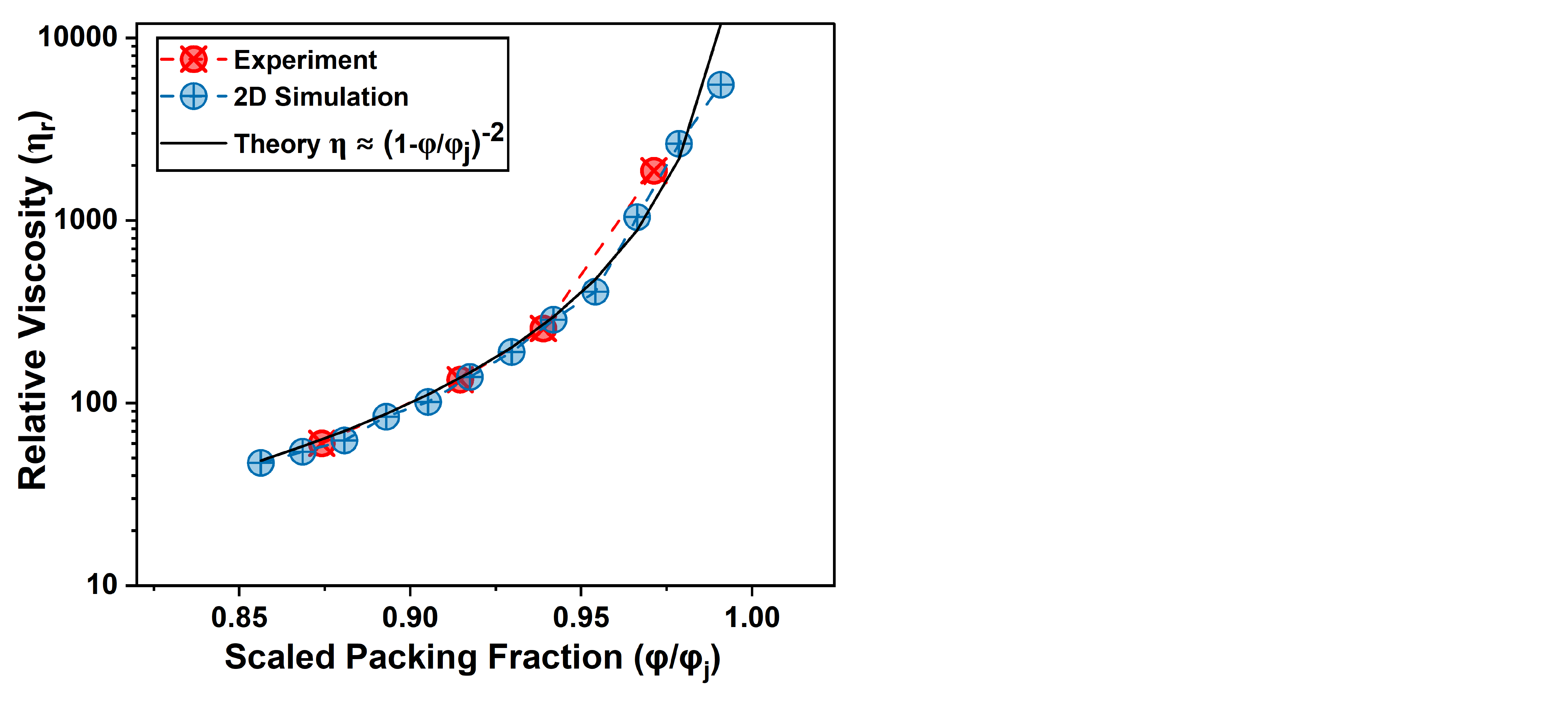}
    \caption{\textcolor{black}{Frictional viscosity $\eta_r(\sigma \to \infty)$ (circles) plotted as a function of $\phi/\phi_J$ and compared with experimental data from Royer et al.~\cite{Royer_2016} (square). The solid line shows $\eta_r(\phi) \sim (1-\phi/\phi_J)^{-2}$.
}}
    \label{fig:s5}
\end{figure}



\clearpage
\section{Training Method}

The first step in predicting the Frictional Contact Network (FCN) within a suspension using Graph Neural Networks (GNN) involves converting the configuration into a graph.
In this representation, each particle is denoted as a node, and an edge connects two nodes if the respective particles are in frictional contact.
In the graph, each edge is associated with edge features ($e_0$) which include the distance between particles ($|r_\text{ij}|$), $x$ and $y$ components of  $\vec{r_\text{ij}}$, and Sine and Cosine of angle $\vec{r_\text{ij}}$ makes with the flow direction.
%
\textcolor{black}{The node features include particle radius as one hot-encoded. One-hot encoding is a commonly used technique in machine learning to represent categorical data as binary vectors. It is especially helpful when dealing with features that have a finite set of distinct categories. In this method, each category is represented by a binary vector, where all elements are 0 except for the position corresponding to the specific category, which is set to 1.}
Then, we subject the node and edge features to a linear transformation that incorporates weight matrices (W(n) and W(e)), bias vectors ($b_n$ and $b_e$), and a hidden dimension ($d_h$) which controls the capacity or expressive power over model \cite{Mandal_2022}.
%
\begin{equation}
    n_0^{'} = W^{(n)} n_0 + b_n
\end{equation}
\begin{equation}
    e_0^{'} = W^{(n)} e_0 + b_n
\end{equation}

This transformation enables our model to map the input features into a higher-dimensional space, empowering the model to learn and represent more intricate patterns and relationships within the dataset.
Next, edge and node features are passed through $N_l$ layers of a residual graph convolutional network (ResGCN) that adds a layer's original input to its output, creating a skip connection. This mitigates information loss during the learning process and enables more effective training of deeper graph neural networks \cite{Guohao_2020, li_2019deepgcns}. 

\begin{equation}
    \begin{split}
        h_v^{l} &= h_v^{l-1} \\
        &\quad + \sum_{u \in N(v)} S \left( W_f^{l}(z_{v, u}^{l-1}) + b_{f}^{l} \right) \odot R \left( W_s^{l}(z_{v, u}^{l-1}) + b_{s}^{l} \right)
        \label{eq:4}
    \end{split}
\end{equation}

Here, $h_v^{(l)}$ is the updated node feature or hidden state for node $v$ at layer $l$, $ \mathcal{N}(v) $ denotes the neighborhood of node $v$, i.e., the set of nodes connected to $v$. 
$f$ is a function that takes as input the features of neighboring nodes, i.e.,  $h_v^{(l-1)}$ and $h_u^{(l-1)}$ and their edge features $e_\text{uv}$, concatenating the features and applying non-linear aggregation function, $\sum_{u \in N(v)}$ is the summation over all neighboring nodes $u$ of node $v$ and $h_v^{(l-1)}$ is the original node feature which will be added to the output of GCN. 
$W_f^l$ and $W_s^l$ are the trainable weight matrices acting on the node and edge features, and $b_f^l$ and $b_s^l$ are the bias vectors. $z_\text{u,v}^\text{l-1}$ is the concatenating vector of the nodes $v$ and $u$ along with their edge feature $e_\text{v,u}$ in the previous layer. $\odot$ represents element-wise multiplication. $S$ and $g$ are
aggregation function $S = {e^{x_i}}/{\sum_j e^{x_j}}$ and activation function, $\text{ReLU}(x) = \max(0, x)$, respectively. We used \href{https://pytorch-geometric.readthedocs.io/en/latest/generated/torch_geometric.nn.models.DeepGCNLayer.html#torch_geometric.nn.models.DeepGCNLayer}{Pytorch Geometric DeepGCN} package to implement our ML.

During training, the model's parameters are adjusted iteratively to align the predicted outputs more closely with the actual targets through the loss function. Traditional cross-entropy (CE), also known as categorical cross-entropy or log loss, is a widely used loss function in machine learning, particularly in classification tasks. 
%
%
In the \href{https://pytorch.org/docs/master/generated/torch.nn.BCEWithLogitsLoss.html#torch.nn.BCEWithLogitsLoss}{Binary Cross Entropy (BCE)} with logits, there are only two possible classes (commonly denoted as 0 and 1), the goal is to predict the probability that an instance belongs to class 1 (the positive class, particles are in contact). In contrast to standard BCE, it merges a Sigmoid function and the conventional BCE into a single class, providing numerical stability compared to using a separate Sigmoid followed by a conventional BCE.
\begin{equation}
    \text{BCE} = - w_n \left[ p_i \log\sigma(q_i) + (1 - p_i) \log(1 - q_i) \right]
\end{equation}

Here, $w_n$ is a manual rescaling weight, $p_i$ denotes the true probability, and $q_i$ is the predicted probability of the instance belonging to class 1 according to the model. This relationship penalizes the model when it predicts a probability close to 0 for the case the actual data belongs to class 1 and vice versa. For instance, in the case of a prediction belonging to class 1 ($ p_i = 1 $), the loss term is $ -\log(q_i) $, while belonging to class 0 ($ p_i = 0 $), the loss term is $ -\log(1 - q_i) $. The major limitation of CE is that it treats all classes equally, which can be problematic in the case of an imbalanced data set and thus can make the model biased toward the majority class. 

To improve model performance and tackle imbalanced datasets, we experimented with various loss functions such as weighted cross entropy, a modified version of CE where we can manually insert a rescaling weight value, and focal loss \cite{lin_2017focal}, allowing us to adjust the contribution of each class by manipulating factors. Although BCE worked well for most predictions, focal loss and weighted cross entropy led to 1-3\% better accuracies in scenarios where the model was trained with $\sigma/\sigma_0=10$ and $\phi = 0.76$. 


Hyperparameters were determined by analyzing loss values for both the training and testing configurations. Ultimately, we selected hyperparameters, $d_h$ set to 64, $N_l$  of 2, and a learning rate ($\lambda$) of 0.005. This systematic approach ensures the optimization of the model's performance based on empirical evidence from the training and testing phases.  
\section{Performance Measurements}

Conventional performance metrics such as accuracy can be deceptive in an imbalanced dataset where one class is much more dominant than the other. A model can reach high accuracy by considering the majority class for all instances, disregarding the minority class completely. 
Several metrics are used to better illustrate a model's performance in an imbalanced dataset \cite{erickson2021magician}.
In addition to average accuracy, we measured average precision measure, recall, F1 score, AUC, and specificity.

Precision assesses the ratio of correctly predicted positive instances to the total number of positive predictions made by the model. Recall, referred to as sensitivity, evaluates the fraction of actual positive cases in the dataset that are correctly identified as positive by the model. metrics focus on the model's performance in the positive class. The F1 score considers precision and recall, making it a more balanced measure for imbalanced datasets. 
It ensures that the model performs well for both the majority and minority classes. AUC evaluates the model's ability to differentiate between positive and negative classes over a range of threshold settings. It is insensitive to class distribution and provides an aggregate measure of model performance. 
Specificity quantifies the model's effectiveness in accurately detecting negative instances from the total of actual negative instances. \cite{erickson2021magician}. 
It complements sensitivity and is especially relevant in imbalanced datasets with the predominant negative class.
Since the F1 score and AUC are relevant metrics for imbalanced datasets, their high values alongside high accuracy reinforce the model's ability to handle the imbalance.
\renewcommand{\thetable}{S\arabic{table}} 
\begin{table}
\caption{Comparison between different metrics to evaluate the performance of 
\textcolor{blue}{the models} \sout{machine learning} at two $\phi$s of 0.76 and 0.80 and $\sigma/\sigma_0$ of 10 to 100}
\label{tab:s1}
\resizebox{\columnwidth}{!}{%
\fontsize{5}{10}\selectfont 
\begin{tabular}{@{}cccccccc@{}}
\textbf{$\sigma/\sigma_0$} &
\textbf{\(\phi\)} &
\rotatebox[origin=c]{90}{\textbf{\parbox{1.5cm}{\centering Average \\[-1ex] Accuracy}}} &
\rotatebox[origin=c]{90}{\textbf{\parbox{1.5cm}{\centering Weighted \\[-1ex] F1-Score}}} &
\rotatebox[origin=c]{90}{\textbf{Precision}} &
\rotatebox[origin=c]{90}{\textbf{Recall}} &
\rotatebox[origin=c]{90}{\textbf{Specificity}} &
\rotatebox[origin=c]{90}{\textbf{AUC-ROC}} \\ \hline
\multirow{2}{*}{\textbf{10}}  & \multicolumn{1}{c|}{\textbf{0.76}} & 0.966 & 0.965 & 0.967 & 0.966 & 0.862 & 0.931 \\
                              & \multicolumn{1}{c|}{\textbf{0.80}} & 0.982 & 0.982 & 0.983 & 0.982 & 0.904 & 0.952 \\
\multirow{2}{*}{\textbf{20}}  & \multicolumn{1}{c|}{\textbf{0.76}} & 0.957 & 0.959 & 0.966 & 0.957 & 0.992 & 0.971 \\
                              & \multicolumn{1}{c|}{\textbf{0.80}} & 0.98  & 0.981 & 0.983 & 0.98  & 0.99  & 0.984 \\
\multirow{2}{*}{\textbf{50}}  & \multicolumn{1}{c|}{\textbf{0.76}} & 0.918 & 0.928 & 0.954 & 0.918 & 0.999 & 0.954 \\
                              & \multicolumn{1}{c|}{\textbf{0.80}} & 0.961 & 0.964 & 0.973 & 0.961 & 0.998 & 0.978 \\
\multirow{2}{*}{\textbf{70}}  & \multicolumn{1}{c|}{\textbf{0.76}} & 0.911 & 0.923 & 0.954 & 0.911 & 0.999 & 0.95  \\
                              & \multicolumn{1}{c|}{\textbf{0.80}} & 0.955 & 0.96  & 0.971 & 0.955 & 0.999 & 0.975 \\
\multirow{2}{*}{\textbf{100}} & \multicolumn{1}{c|}{\textbf{0.76}} & 0.907 & 0.92  & 0.955 & 0.907 & 1     & 0.949 \\
                              & \multicolumn{1}{c|}{\textbf{0.80}} & 0.951 & 0.956 & 0.97  & 0.951 & 1     & 0.973
\end{tabular}%
}
\end{table}

According to the results obtained for our model trained at packing fractions of $\phi$ of 0.76 and 0.80 and stress $\sigma/\sigma_0=10$ (Table \ref{tab:s1}), all the metrics, including F1 score and AUC, are high and close to the value of average accuracy indicating the robustness of our model to capture both classes.
For example, the prediction results for $\phi=0.8$ and $\sigma/\sigma_0=50$ show a high precision value, meaning 97.3\% of the time the model predicts something as positive, it's positive; a high Recall value indicating the model catches 96.1\% of the actual positive cases; an exceptionally high specificity, meaning the model rarely identifies negative cases as positive (only 0.2\% misclassification); a very good F1 score (96.4\%), indicating a high balance between precision and recall; and a high AUC value (97.8\%), meaning the model can effectively distinguish between positive and negative cases

\begin{figure*}
\centering
    \includegraphics[trim = 0mm 280mm 320mm 0mm, clip,width=0.95\textwidth,page=11]{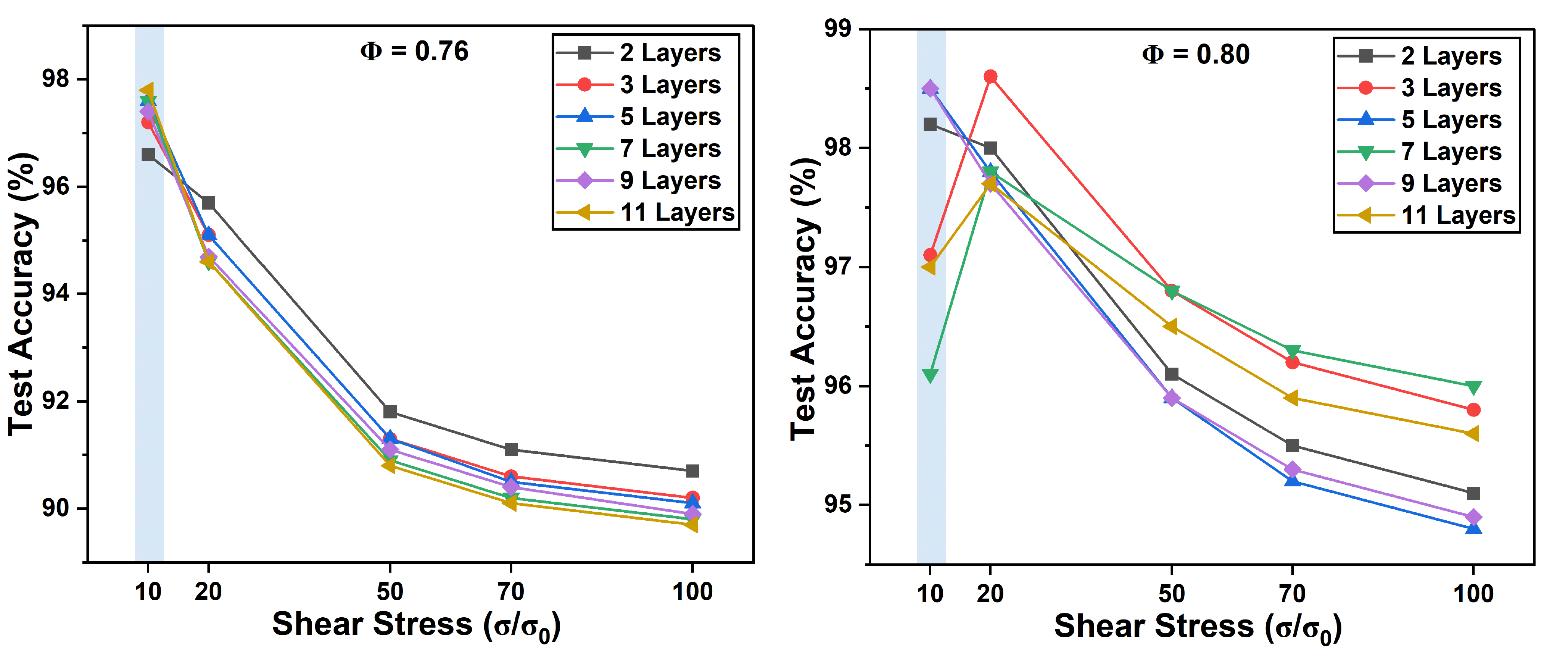}
    \caption{
    Test accuracy as a function of $\sigma/\sigma_0$ for  $\phi=$ 0.76 (left) and 0.80 (right) while all have been trained at $\sigma/\sigma_0=10$ with different number of layers of DeepGCN.
    }
    \label{fig:s6}
\end{figure*}

\paragraph{Obtaining optimum number of layers:}
When the model is trained with an increased number of layers, the process of message-passing delves deeper into the network, extracting information from nodes that extend beyond its immediate neighbors. This augmentation proves particularly beneficial in scenarios where obtaining extensive information is crucial for enhancing predictive accuracy, especially when considering the long-range interactions among particles, which can significantly influence the FCN and overall rheological response. 
Figure~\ref{fig:s6} illustrates the model's efficacy under initial conditions with values of $\phi$ = 0.76 and 0.80 as a function of $\sigma/\sigma_0$ with different numbers of deepGCN layers, showcasing the performance across varying layer depths. 
Notably, even with a shallow model architecture featuring just two layers, the prediction is highly accurate and emphasizes the robustness of the model's capabilities. Scientifically, this highlights the model's adeptness in capturing intricate relationships and dependencies within the data. Regarding the lower accuracies for $\phi$ = 0.76, as the number of deepGCN layers are varied from 2 to 11 without a significant impact on the accuracy. This suggests that factors beyond the number of layers might be influencing performance, such as adjustments to other training parameters. 
Beyond a certain depth (2 layers) the performance of the GNN deteriorates regardless of the parameter $\phi$ it's trained with. This could indicate an issue with overfitting or vanishing gradients as the network becomes deeper. This suggests that as the GNN processes messages over layers, the node representations become increasingly similar to their neighborhood information. This could lead to a loss of discriminative power, causing the model's generalization performance to worsen.

\paragraph{Structure Factor Predictions}
The structure factor predictions of suspensions under different conditions are visualized for particles in the frictional contact network:
\begin{equation}
    S(q) = \frac{1}{N_c} \sum_{i=1}^{N_c} \sum_{j=1}^{N_c} \exp(-i \mathbf{q} \cdot (\mathbf{r}_i - \mathbf{r}_j)),
\end{equation}
where $S(q)$ is the structure factor as a function of the wave vector q, $N_c$ represents the total number of particles being part of FCN in the system, $\sum_{i=1}^{N_c} \sum_{j=1}^{N_c}$ are double summations over all pairs of particles in the system, $\mathbf{r}_i - \mathbf{r}_j$ the vector difference between the positions of particles $i$ and $j$. This form, $ln|S(q_x, q_y)|^2$ of the structure factor, enhanced clarity and provided a more visually appealing representation. Additionally, only 1/4 of the entire structure factor was considered to reduce computational generation time.

\paragraph{Training at different stress levels:}
In the main text, we presented results when the model is trained for a stress $\sigma/\sigma_0=10$ (Fig. 6 on the main text), as well as prediction accuracy while training at $\sigma/\sigma_0=5, 10, 50$. Figure~\ref{fig:s7} complements the accuracy calculations with the visual representation and structural factor for $\phi = 0.76$ and 0.80. 
\paragraph{Predictions for different volumetric mixing ratio $\alpha$:}
Figure~\ref{fig:s8} illustrates the prediction of the structure factor of FCN and absolute error of structure factor for systems with particle size ratio $\Delta=1.4$, and various values of the volumetric mixing ratio with $\alpha = [0.1, 0.3, 0.5, 0.7, 0.9]$ when it has been trained only at $\alpha$ = 0.5. In the main text, we have also demonstrated that the model can accurately capture the FCN when trained at $\alpha$ = 0.1 or $\alpha$ = 0.9. 
%
\paragraph{Predictions for different size ratios $\Delta$:}
Figure~\ref{fig:s9} displays examples of the prediction of FCN for systems consisting of $N=400$ and 2000 particles at various values of particle size ratios ($\Delta$) from 2 to 6 (for a constant $\alpha=0.5$) when the model is only trained at configuration with $N=400$ particles and $\Delta$ = 1.4. The rows from top to bottom exhibit particles in contact, frictional network, structure factor of predicted the FCN and absolute error of structure factor. We observe a good accuracy based on two observations (i) very minimal red particles (red particles show instances where the contact is not correctly predicted), and (ii) low absolute error in predicting the structure factor. 
\begin{figure*}
    \includegraphics[trim = 0mm 0mm 235mm 0mm, clip,width=0.95\textwidth,page=12]{images/paper1_v3.pdf}
    \caption{
    Structure factor predictions based on FCN along with the corresponding network for two different fixed values of $\phi$. It depicts predictions for various $\sigma/\sigma_0$s when the models are only trained at the shear stress $\sigma/\sigma_0=10$. The top row indicates the prediction at various stresses at $\phi$ = 0.76; the bottom row is the same but for $\phi$ = 0.80.
    }
    \label{fig:s7}
\end{figure*}

\begin{figure*}
    \includegraphics[trim = 0mm 260mm 200mm 0mm, clip,width=0.95\textwidth,page=13]{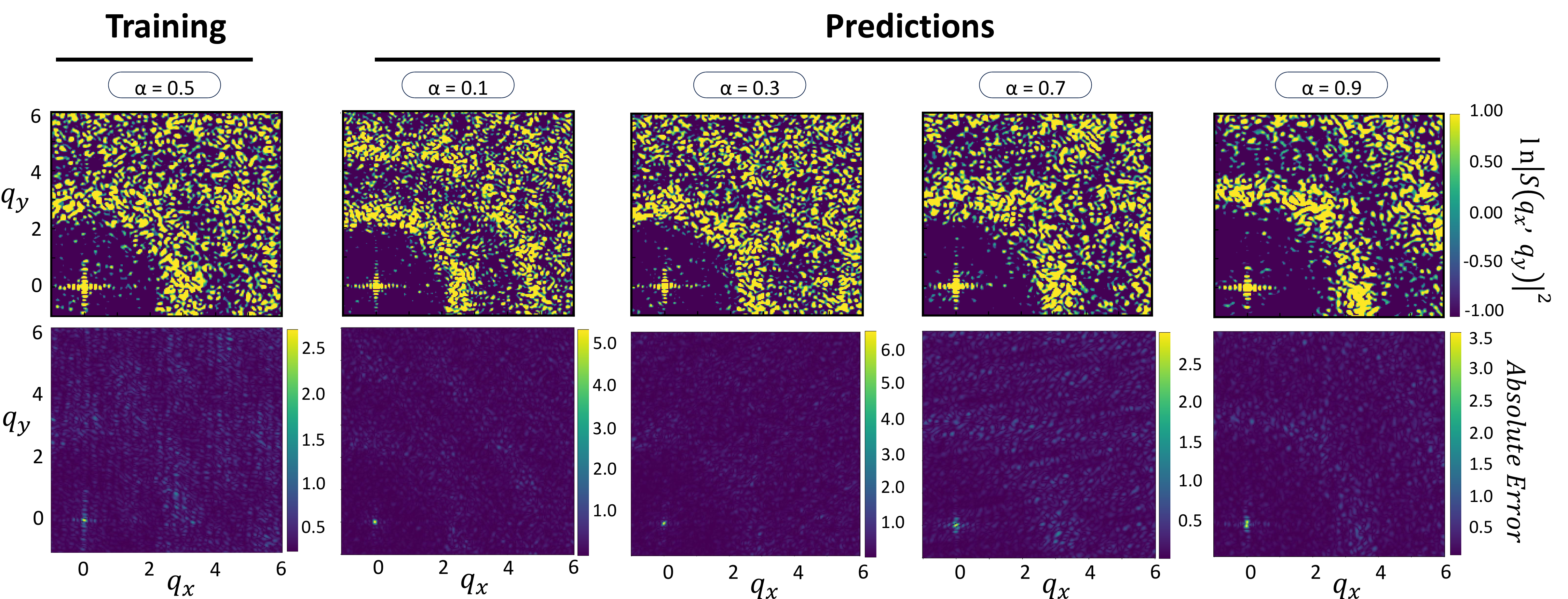}
    \caption{
    Structure factor predictions for systems with various volumetric mixing ratios ($\alpha$) from 0.1 to 0.9 for $\Delta=1.4$ when the model has been trained only at $\alpha$ = 0.5. The top row visualizes the structure factor of predicted FCN and the bottom shows the absolute error of the actual and predicted structure factor.
    }
    \label{fig:s8}
\end{figure*}

\begin{figure*}
    \includegraphics[trim = 0mm 5mm 0mm 0mm, clip,width=0.95\textwidth,page=14]{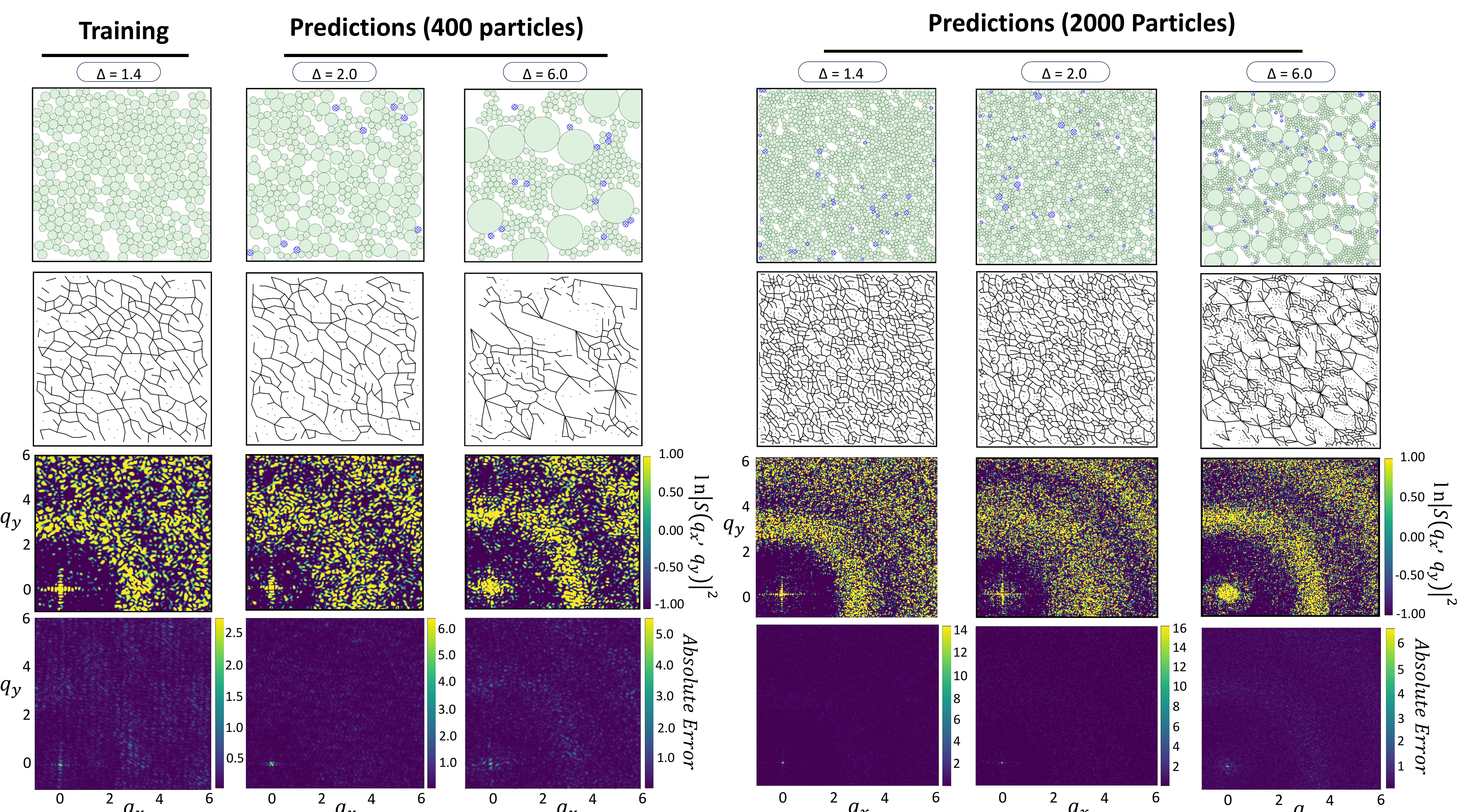}
    \caption{
    Frictional contact predictions for systems consisting of $N=400$ and $N=2000$ particles at various particle size ratios ($\Delta$) from 2 to 6 (for a constant $\alpha=0.5$), when the model is only trained at $\Delta$ = 1.4 and $N=400$. The rows from top to bottom depict particles in contact, the frictional network, the structure factor of the predicted FCN, and the absolute error of the structure factor. Misidentified particles are shown in \textcolor{black}{lattice-patterned blue}.
    }
    \label{fig:s9}
\end{figure*}

\pagebreak
\newpage
\end{appendix}
\end{widetext}
\end{document}